\documentclass[prd,nofootinbib,floats,superscriptaddress,eqsecnum,tightenlines,11pt]{revtex4}

\usepackage{hyperref}
\usepackage{graphicx}
\usepackage{amsmath,amssymb,amsfonts,amsthm,latexsym,stmaryrd}
\usepackage{marginnote}
\usepackage{color}

\def\be{\begin{equation}}
\def\ee{\end{equation}}
\def\ba{\begin{eqnarray}}
\def\ea{\end{eqnarray}}
\def\bas{\begin{subequations}\begin{eqnarray}}
\def\eas{\end{eqnarray}\end{subequations}}

\def\lp{\ell_\text{Pl}}

\def\Re{\text{Re}}

\def\SU{\text{SU}}

\begin{document}

\title{Analytic Continuation of  Black Hole Entropy\\ 
in Loop Quantum Gravity}

\author{Ben Achour Jibril}
\email{benachou@apc.univ-paris7.fr}
\affiliation{Laboratoire APC -- Astroparticule et Cosmologie, Universit\'e Paris Diderot Paris 7, 75013 Paris, France}
\author{Amaury Mouchet}
\email{mouchet@lmpt.univ-tours.fr}
\affiliation{ Laboratoire de Math\'ematiques et Physique Th\'eorique,
\textsc{\textsc{cnrs (umr 7350)}},
F\'ed\'eration Denis Poisson,Universit\'e Fran\c{c}ois Rabelais de Tours
--- \textsc{\textsc{cnrs (umr 7350)}},
F\'ed\'eration Denis Poisson,
Parc de Grandmont 37200 France}
\author{Karim Noui}
\email{karim.noui@lmpt.univ-tours.fr}
\affiliation{ Laboratoire de Math\'ematiques et Physique Th\'eorique,
\textsc{\textsc{cnrs (umr 7350)}},
F\'ed\'eration Denis Poisson,Universit\'e Fran\c{c}ois Rabelais de Tours
--- \textsc{\textsc{cnrs (umr 7350)}},
F\'ed\'eration Denis Poisson,
Parc de Grandmont 37200 France}
\affiliation{Laboratoire APC -- Astroparticule et Cosmologie, Universit\'e Paris Diderot Paris 7, 75013 Paris, France}

\begin{abstract}
We define the analytic continuation of the number of black hole microstates in Loop Quantum Gravity to complex values of the Barbero-Immirzi parameter $\gamma$.
This construction deeply relies on the link between black holes and Chern-Simons theory. Technically, the key point consists in writing the number of microstates as an  integral in
 the complex plane of a holomorphic function, and to make use of complex analysis techniques to perform the analytic continuation.
Then, we study the thermodynamical properties of the corresponding system (the black hole is viewed as a gas of indistinguishable punctures)
in the framework of the grand canonical ensemble where the energy is defined \`a la Frodden-Gosh-Perez from the point of view of 
an observer located close to the horizon. 
The semi-classical limit occurs at the Unruh temperature $T_U$ associated to this local observer. When $\gamma=\pm i$, the entropy reproduces at the semi-classical limit
the area law with quantum corrections. Furthermore, the quantum corrections are logarithmic provided that the chemical potential is fixed to the simple value $\mu=2T_U$.
\end{abstract}

\maketitle

\section{Introduction}
Proposing a  consistent microscopic explanation of  the celebrated thermodynamical properties of black holes found by Bekenstein \cite{Bekenstein} and Hawking \cite{Hawking} in the 70's
is certainly one of the most important theoretical tests for any candidates to  quantum gravity. It is indeed expected from any theories of quantum gravity to provide a framework for
understanding the  statistical physics behind these thermodynamical properties and, more particularly, for finding the fundamental excitations responsible for the black hole entropy. 
Almost four decades  after the discovery by Bekenstein and Hawking, in different approaches of quantum gravity, we have witnessed a lot of progresses in the field of black hole
thermodynamics. There exist  dozens of different technical derivations of the black hole entropy from semi-classical arguments or from theories of quantum gravity. However, it is
fair to observe that the question of the microscopic structure of black holes is not totally resolved.  We still do not  precisely know what does the huge black hole entropy really
count: the counting is clear but the content is not. Even in the much simpler case of three dimensional gravity, the nature of the fundamental excitations of the celebrated  BTZ black hole  \cite{BTZ1,BTZ2} responsible for its entropy remains 
unknown (see  \cite{Carlip} for a review). From the point of view of Loop Quantum Gravity, the derivation of black hole entropy relies mainly on the idea that this entropy
counts essentially the number of ways  the macroscopic horizon area can be obtained as a sum of fundamental excitations. The area law has been recovered 
originally for non-rotating macroscopic black holes in \cite{Rovelli,Ashtekar} and then the counting was upgraded in different ways \cite{BHentropy3,BHentropy4,BHentropy6,BHentropy7}.

\subsection*{Black Hole entropy with a real $\gamma$}
In fact, the derivation of the black hole entropy in Loop Quantum Gravity relies essentially on two ingredients.  First, the black hole is defined by its horizon which is assumed to be  a boundary in space-time. Second, the classical 
geometrical properties of the horizon (null-surface, no expansion etc...) are turned into quantum constraints which make the quantum black hole degrees of freedom to be described by a Chern-Simons theory. More precisely, the symplectic structure
of the black hole degrees of freedom are given by the Chern-Simons Poisson bracket on a punctured two-sphere $S_2$ with gauge group $G=SU(2)$ and the level $k$ is proportional to the horizon area $a_H$ \cite{ENP}. 
The compact  group $SU(2)$ is reminiscent of the internal gauge group of gravity expressed in terms of the Ashtekar-Barbero connection.  Regarding the punctures, there is a priori an arbitrary number of them, denoted $n$
in the sequel. They are the fundamental excitations, associated to a spin-network in the space manifold and colored with $SU(2)$ unitary irreducible representations $(j_1,\cdots,j_n)$, which cross the horizon. 
The macroscopic horizon area 
\begin{eqnarray}\label{area spectrum}
a_H=\sum_{\ell=1}^n a_\ell \;\;\;\text{with}\;\;\; a_\ell=8\pi \ell_P^2 \gamma \sqrt{j_\ell(j_\ell+1)}
\end{eqnarray}
 results from the sum of the fundamental excitations  $a_\ell$ carried by the punctures. Here $\gamma$ denotes the
Barbero-Immirzi parameter which enters in the spectrum of the area operator in Loop Quantum Gravity and $\ell_P=\sqrt{G\hbar}$ is the Planck length \cite{Barbero,Immirzi}. Therefore, in this picture, the black holes micro-states are elements of the physical Hilbert space of the
$SU(2)$ Chern-Simons theory on a punctured two-sphere, and the entropy  of a black hole whose horizon has an area $a_H$ counts the number of such micro-states provided that the condition 
(\ref{area spectrum}) is fulfilled.  

Then, one needs to understand the Hamiltonian quantization of Chern-Simons theory. 
For this reason, the three dimensional space-time required for the Chern-Simons theory is assumed to be locally of the form $\Sigma \times \mathbb [0,1]$ where $\Sigma$ is a Riemann surface. 
When the gauge group $G$ is compact, the (canonical and covariant) quantization 
is very well understood. It has been studied intensively since the discoveries by Witten that first three dimensional gravity can be reformulated in terms of a Chern-Simons theory \cite{WittenCS}, and then that the quantum
amplitudes of the Chern-Simons theory are closely related to topological three-manifolds invariants and knots invariants \cite{WittenJones}. The so-called combinatorial quantization is certainly one of the most powerful 
Hamiltonian quantization scheme of the Chern-Simons theory because it can be applied not only to a large class of compact \cite{CQ1,CQ2,CQ3} and non-compact gauge groups \cite{BNR,Cath1,Cath2,Cath3,Cath4} but also to any 
punctured Riemann surfaces $\Sigma$ (Note that Loop Quantum Gravity techniques are very successful when there is no cosmological constant \cite{NP1,NP2}, and the case of non-vanishing cosmological is still under construction
\cite{Pranz1,Pranz2}).
When the gauge group is $G=SU(2)$ and the surface $\Sigma$ is a punctured two-sphere, the physical Hilbert space ${\cal H}(j_\ell)$, where $(j_1,\cdots,j_n)$ color the punctures, has a
 very simple structure. It can be constructed from the representation theory of
the quantum group  $U_q(\mathfrak{su}(2))$ where the quantum deformation parameter $q$ is related to the level $k$ of the Chern-Simons theory by $q=\exp(i\pi/(k+2))$. Because the gauge group is compact, $k$ is necessarily 
integer and then $q$ is a root of unity. This result has important consequences: it makes the number of unitary irreducible representations (irreps) of  $U_q(\mathfrak{su}(2))$ finite and is somehow responsible for the good convergence properties of the Chern-Simons quantum amplitudes. As for the classical Lie group $SU(2)$,  the irreps are labeled by spins $j$ (which are half-integers), they are realized on a  finite dimensional vector space $V_j$ of dimension $d=2j+1$, but  
contrary to the classical case, $j$ is bounded from above according to the inequality $j \leq k/2$. Finally, the physical Hilbert space ${\cal H}_k(j_\ell)$ is  the space of $U_q(\mathfrak{su}(2))$ invariant vectors (equivalently intertwiners) in the
tensor product of the representations vector spaces $V_j$  
\begin{eqnarray}
{\cal H}_k(j_\ell) = \lbrace \text{Inv} ( \bigotimes_{\ell=1}^n V_{j_\ell}  ) ;  d\mu \rbrace
\end{eqnarray}
where $d\mu$ is inherited from the (right or left invariant) Haar measure on the quantum group $SU_q(2)$ (the standard notation for the space of polynomial functions or its Cauchy completion).  
In that respect, the black hole microstates from the point of view of Loop Quantum Gravity are
intertwiners between the representations coloring the links crossing the horizon. These representations are viewed as $U_q(\mathfrak{su}(2))$ representations.
Obviously, ${\cal H}_k(j_\ell)$ has a finite dimension given by \cite{BHentropy7}
\begin{eqnarray}\label{dimCS}
{\cal N}_k(d_\ell) = \frac{2}{k+2} \sum_{d=1}^{k+1} \sin^2(\frac{\pi d}{k+2}) \prod_{\ell=1}^n \frac{\sin(\frac{\pi}{k+2}d d_\ell) }{\sin(\frac{\pi}{k+2}d) }
\end{eqnarray}
where $d_\ell=2j_\ell +1$. This formula (or its classical limit when $k \rightarrow \infty$) is known as the Verlinde formula and is  in the core of the calculation of the black hole entropy in Loop Quantum Gravity.
The counting of microstates leads to the Bekenstein-Hawking area law provided that the Barbero-Immirzi parameter $\gamma$ is fixed to a particular finite and
real value. Recovering the area law is clearly a non-trivial success of Loop Quantum Gravity but the important  role played by $\gamma$ in this result, whereas it is totally irrelevant in the
classical theory, has raised important questions on the validity of this counting.  

\subsection*{Black Hole entropy with  $\gamma=\pm i$}
These last couple of years may have brought new interesting insights to resolve this puzzle. These new ideas mainly relies on an interpretation of $\gamma$ as a regulator in the theory and not as a fundamental
constant. Indeed, the Ashtekar-Barbero connection together with the Barbero-Immirzi parameter $\gamma$ has been introduced \cite{Barbero,Immirzi} in Loop Quantum Gravity to circumvent the problem of the reality conditions.
Therefore, considering a real parameter $\gamma$ appears as a technical trick (a kind of Wick rotation) to avoid having to work with the original complex Ashtekar variables and to start the quantization, at least at the kinematical level. 
So far, no one knows how to start the quantization of gravity in the complex Ashtekar formulation, which corresponds to taking $\gamma=\pm i$.
In that sense, the introduction of a real $\gamma$ in the theory is a necessity in Loop Quantum Gravity, exactly as it is necessary to consider a Wick rotation in quantum field theory to compute 
path integrals for instance. However, we should return to the complex value $\gamma=\pm i$ at some point. Furthermore, this requirement is  strongly  suggested by the classical theory.
Indeed, it has been known for a long time that  the Ashtekar-Barbero connection is not a full space-time connection whereas the original complex connection is certainly one \cite{Samuel,Alexandrov,Geiller1}. This observation has led to the idea
that, even though the Ashtekar-Barbero connection is suitable for the kinematical quantization of Loop Quantum Gravity, it could not be the proper variable to deal with the dynamics and the Hamiltonian constraint. This fact was concretely realized 
in three space-time dimensions where, after introducing a Barbero-Immirzi parameter $\gamma$ in the model \cite{Geiller2,Geiller3}, it has been shown that  the Ashtekar-Barbero connection could not lead to a resolution of the Hamiltonian constraint,
and it is necessary to work in the self-dual variables to solve the dynamics \cite{Jibril1,Jibril2}. Moreover, we showed that if the spectrum of the geometric operators is discrete and $\gamma$-dependent at the kinematical level,
it becomes continuous and $\gamma$-independent at the physical level. Formally, it is possible to recover the physical spectrum from the kinematical one sending $\gamma$ to the values $\pm i$ and, at the same time,
sending the representations (which label the edges of the spin-networks) $j_\ell$ to the complex values $(-1 + is_\ell)/2$ where $s_\ell$ is a real number. Considering $j_\ell=(-1 + is_\ell)/2$ instead of $j_\ell$ half-integer is easily 
interpreted as coloring  the spin-networks 
in terms of $SU(1,1)$ irreps (in the continuous series) instead of $SU(2)$ irreps. It is interesting to notice that a simple analytic continuation of the physical parameter reproduce the exact physical expressions for the eigenvalues of 
geometric operators.  All these strongly suggests that one should interpret $\gamma$ as a regulator  which should be sent to the complex value $\gamma=\pm i$.

Another strong indication comes  from the physics of black holes in Loop Quantum Gravity.  It was shown in \cite{FGNP}, that for a fixed number of punctures and in the limit of large spins, the analytically-continued dimension of the 
$\SU(2)$ Chern-Simons Hilbert space (which can be understood as the weight assigned to a black hole microscopic configuration)  behaves exactly as $\exp(a_H/4\lp^2)$. The analytic continuation consists in first replacing the discrete dimensions 
$d_\ell$ in the formula (\ref{dimCS}) by the complex numbers $(-1 + is_\ell)/2$ and then in fixing $\gamma$ to the value $\pm i$ in order for the area spectrum (\ref{area spectrum}) to remain positive and real. Interestingly, it was observed that the same 
analytically-continued technique can be naturally applied to compute the three-dimensional BTZ black hole entropy \cite{BTZ}. In this case however, it is the sign of the cosmological constant and not $\gamma$ that plays the role of a regulator.
The proposal of \cite{FGNP} is supported by remarkable additional facts \cite{KMS,CMC,BN,Yasha1, Yasha2,Yasha3,radiation,Muxin}.  Finally, all these observations strongly suggest that we should take seriously the idea to return in some way 
to the original self-dual variables in Loop Quantum Gravity. Either we extend the analytic continuation method to the full kinematical Hilbert space and eventually to the Spin-Foam amplitudes, or we use this analytic continuation process to guide
us towards a resolution of the reality conditions. In both cases, it is first of all necessary to understand in a more (mathematically) rigorous way the analytic continuation proposed in \cite{FGNP} which is precisely the aim of this article. 

 \subsection*{Using complex analysis}
What has been proposed in \cite{FGNP} is an analytic continuation of the dimension (\ref{dimCS}) viewed as a function on a discrete set\footnote{The status of the Chern-Simons level $k$ is not clear in this process \cite{FGNP}. It seems that
$k$ remains real. Nonetheless, it is possible to assume that $k$ is sent to a  purely imaginary variable, exactly as the summation variable $d$, which makes the sum (\ref{dimCS}) unchanged (at the classical limit when $k$ is large). The role of $k$ will
be totally clarified in the present article.} where the discrete variables $d_\ell \in \mathbb N$ 
becomes complex (more precisely, purely imaginary). Schematically, the analytic continuation is defined by the map $d_\ell \rightarrow is_\ell \in i \mathbb R$.
 It goes without saying that such a process is highly non-trivial and in general not unique. However, complex analysis provides
us with powerful tools to define rigorously this kind of transformations. These tools (essentially Morse theory) have been extensively used by Witten \cite{Witten analytic} to study the analytic continuation of Chern-Simons theory from integer values of
$k$  to complex values. As we want to perform an analytic continuation of the dimension of the Chern-Simons Hilbert space, the methods developed in \cite{Witten analytic} seem totally appropriate for our purposes. We show this is indeed the case.
Not only we obtain a mathematically well-defined analytic continuation of (\ref{dimCS}) (with ambiguities related to choices of the contour in the complex plane) but we also compute its (semi-classical) asymptotic behavior.
Interestingly, we obtain the result of \cite{FGNP} with some corrections. 

To obtain these results, we proceed in different steps. First we 
write (\ref{dimCS}) as an integral in the complex plane because we interpret the sum over $d$ in (\ref{dimCS}) as a sum of residues. Hence, 
(\ref{dimCS}) can be written as $\int_{\cal C} dz \, \mu(z) F(z)$ where $\cal C$ is a closed contour in the complex plane,
$\mu$ a measure independent of the colors and $F(z)$ a holomorphic function which depends on the colors. Then, we perform an analytic continuation
to a purely imaginary level $k \rightarrow i \lambda$ ($\lambda \in \mathbb R^+$) keeping the contour $\cal C$ fixed. We see immediately that such a continuation 
leads to a vanishing integral and then to a non-consistent analytic continuation of the number of microstates. However, when the dimensions $d_\ell$ become
are non longer integer (they are purely imaginary dimension $d_\ell \rightarrow i s_\ell$, $s_\ell \in \mathbb R^+$, in our case), then the analytic continuation
is non-trivial and leads to a very interesting analytic continuation of (\ref{dimCS}). In fact,  this continuation is uniquely defined up to a discrete ambiguity which
 is very similar to the one raised by Witten \cite{Witten analytic} (and mainly based on \cite{Berry}) when he recalls the construction of the analytic continuation of the Bessel function. 
More precisely, the contour is defined up to a translation $z\mapsto z+ ip2\pi$, $p\in \mathbb N$, in the complex plane.
Fortunately, there exists one natural choice of the contour which leads to the expected semi-classical behavior of the analytic continued number of microstates. 
Computing such an asymptotic  is technically rather involved. To simplify the problem, we first assume that the horizon is
punctured by $n$ edges all colored with the same complex spin $d=is$: this defines what is called  the one color black hole model in the article.
In that case, the expression of the function $F(z)$ introduced above simplifies and the complex integral  can be  reformulated as an
 integral$\int_{\cal C} dz \, \mu(z) \exp(nS(z)) $ where $n$ plays the role of the inverse Planck constant and the action $S(z)$ is a holomorphic 
function (which depends on $s$) in the complex plane ($z \in \mathbb C$). As shown in \cite{Amit}, $n$ becomes large at the semi-classical limit.
As a consequence, the semi-classical behavior is  determined from the saddle point approximation, hence from the analysis of the critical points of the action $S(z)$. 
We show that there is one critical point which dominates the classical limit. Following the saddle point approximation techniques at the vicinity of this critical point,  
we obtain  the large $n$ behavior of the number of microstates. This enables us to obtain the microcanonical entropy for the one color black hole model
which reproduces the area law supplemented with quantum corrections.
We go further and compute the canonical and grand canonical entropy. Interestingly, with a simple choice of the chemical potential, the quantum corrections are logarithmic, of the form
$-3/2 \log(a_H/\ell_p^2)$, i.e. the same as the one obtained for black holes with a real $\gamma$.
We reproduce the analysis in the case where the punctures are colored by an arbitrary but finite number $p$ of different colors $(s_1,\cdots,s_p)$ assuming that each number $n_p$ of punctures colored by $s_p$ becomes large
in the same way at the semi-classical limit. We show that the black hole entropy still satisfies the area law in this general case with  quantum corrections (which are logarithmic with a particular choice of chemical potential). Our 
calculation is  rigorous and gives a mathematical justification of \cite{FGNP}. In that respect, it is particularly interesting to note that the ``naive" analytic continuation presented in \cite{FGNP} reproduces, at the semi-classical 
limit, the good leading-order term. Furthermore, the method we are developing in this article can be applied to generic situations and we hope to extend it to construct Spin-Foam models or kinematical scalar products for complex Ashtekar
gravity.

\subsection*{Organization of the article}
The article is organized as follows. After this  introduction, we propose in Section II  an overlook of  the recent results obtained in the context of complex black holes. 
First, we  recall  the main results of the ``naive" analytic continuation
in \cite{FGNP} and show how the area law is immediately recovered at the semi-classical limit. Then, we recall and adapt the analysis of the black hole (canonical and grand canonical) partition function done in \cite{Amit} when the area spectrum
is continuous. This analysis is particularly important for our purposes because the study of the  thermodynamical limit  enables us to define the semi-classical limit as the regime where the horizon area $a_H$ is large 
together with the number of punctures $n$ and their colors $s_\ell$. Section III is devoted to studying the analytic continuation of (\ref{dimCS}) from 
rigorous complex analysis. We start by studying the effect of turning the (discrete) level $k$  into a pure imaginary number $i\lambda$. This is done by expressing the Chern-Simons dimension as an integral $I$
in the complex plane. We show that, if $k$ becomes purely imaginary, the representations $d_\ell$ must no longer be discrete for the analytic continuation to be non-vanishing, hence we replace the integers $d_\ell$ by 
purely imaginary numbers, i.e. $d_\ell=is_\ell$. 
Then, we analyze the asymptotic of the analytic continued version of the number of microstates.
To do so, we first study the case where the $n$ punctures are colored with the same $s_\ell=s$: this defines the one color black hole model. In that case, the integral $I$ can be written
as $I=\int_{\cal C} d\mu(z) \exp(nS(z)) $ where $\cal C$ is the integration contour,  $\int d\mu(z)$ is a well-defined measure and the action $S(z)$ is a holomorphic function which depends on
the color $s$. The saddle point approximation immediately leads us to the large $n$ asymptotic of $I$ and reproduces the area law for the microcanonical entropy. We perform the grand canonical
analysis of the model and show that, with a simple choice of chemical potential, the quantum correction of the entropy are logarithmic.  
We generalize the study to a system of $n$ punctures colored by a finite number $p$ of colors $s_1,\cdots,s_p$. We conclude in Section IV with a brief summary and a discussion. 
We argue on the possibility to adapt these techniques to construct Spin-Foam amplitudes or kinematical scalar products in the self-dual sector of Loop Quantum Gravity.

\section{Complex black holes in Loop Quantum Gravity}
The derivation of black hole entropy relies essentially on the idea that the macroscopic horizon area is realized
as the sum of microscopic contributions (quanta of area) excited by the spin network edges which puncture the horizon. In this picture, the counting of the microstates
leads to the Bekenstein-Hawking area law provided that $\gamma$ is fixed to a particular finite and real value. The fact that this parameter seems to play such a crucial role in the quantum
theory even though it is totally irrelevant at the classical level has raised some doubts. 
New insights have been developed the last years to resolve this contradiction. The first one is that the  quantum degrees of freedom of a spherical black hole  in Loop Quantum Gravity are those
of an $SU(2)$ Chern-Simons theory where the level $k$ is proportional to the horizon area $a_H$  according to the relation $a_H/\ell_p^2=2\gamma (1-\gamma^2) k$ \cite{ENP}.  The second one is 
to consider a black hole with a complex Barbero-Immirzi parameter. This makes totally sense because considering $\gamma$ complex is directly related to considering the analytic continuation of 
Chern-Simons theory to complex values of the level k \cite{Witten analytic}. This last idea was realized concretely in \cite{FGNP} and led in a very direct way to the area law for the black hole entropy
when $\gamma$ takes the special values $\pm i$. This strongly indicates that $\gamma$ is a regulator which should be fixed to $\pm i$ at some point. It corresponds to returning to the original
complex Ashtekar variables. 

This Section aims at recalling first the analytic continuation performed in \cite{FGNP}. Then, we will recall and adapt  (to continuous spins) the  thermodynamical properties of 
the complex black holes studied recently in \cite{Amit}. This enables to characterize the semi-classical regime and, in particular, to show that the number of punctures becomes
large at the semi-classical limit (or equivalently the thermodynamical limit) when $a_H$ is large. 

\subsection{Black holes with $\gamma=\pm i$}
\label{gamma = i}
The computation of the black hole entropy in Loop Quantum Gravity is roughly based on  two formulae. The first one is the expression of the horizon area $a_H$ recalled in 
the introduction (\ref{area spectrum}). 
 The second one is the expression (\ref{dimCS})  of the number of microstates with the level $k$ fixed by the relation $a_H/\ell_p^2=2\pi \gamma (1-\gamma^2) k$ as shown in \cite{ENP}.

For a configuration where $d_\ell=d$ for every $\ell$, it is easy to show that (\ref{dimCS}) reproduces the area law at the semi-classical limit ($d \gg 1$ and $k\gg 1$) 
\begin{eqnarray}\label{discrete asymp}
\log ({\cal N}_k(d_\ell)) \; = \; n \log(d) + o(n) \; = \; \frac{a_H}{4\ell_p^2} \frac{\log(d)}{\pi \gamma d} + o(a_H)
\end{eqnarray}
provided that $\gamma$ is fixed according to $\pi \gamma d = \log(d)$. A similar result holds when we take into account all the possible configurations with 
a different real value for the Barbero-Immirzi parameter. In the latter case, one recovers the area law only if one  assumes the distinguishability of the punctures.
In \cite{FGNP}, we asked the question whether it could make sense to extend the framework of black hole thermodynamics in Loop Quantum Gravity to complex values
of $\gamma$. In fact, we are interested only in the case $\gamma = \pm i$. For the answer to be positive, it is necessary to keep the area $a_H$ (\ref{area spectrum}) real while
$\gamma$ becomes complex. This condition is satisfied only if $j_\ell=(-1 + is_\ell)/2$ or equivalently if the dimension $d_\ell=2j_\ell +1=is_\ell$ becomes purely imaginary.
In that case, $a_H$ is necessarily real and we can easily make it positive by choosing the suitable square root $\sqrt{j_\ell(j_\ell+1)}=\sqrt{-(s_\ell^2 + 1/4)}$ in 
the area spectrum (\ref{area spectrum}).

Concerning the number of states, its analytic continuation is much less obvious to define. The reason is  that  the level $k$ which appear in the upper bound of the sum (\ref{dimCS}) 
becomes purely imaginary $k=i\lambda$. We can suppose $\lambda >0$ without loss of generality. Therefore, the sum becomes a priori meaningless. To make it well-defined, a ``naive" idea is to assume that the label $d$ becomes also purely imaginary,
which would correspond to a discrete analog of turning the integration path (in the integral of a complex-valued function) in the complex plane from the real line $\mathbb R$ to the purely 
imaginary line $i\mathbb R$. With this ``naive" definition of the analytic continuation, it is immediate to show that the number of microstates behaves as
\begin{eqnarray}\label{asymp naive}
\varepsilon^2 \prod_{\ell=1}^n \frac{\sinh(2 \pi s_\ell)}{\varepsilon} =\frac{ \varepsilon^{2-n}}{2^n} \exp(\frac{a_H}{4\ell_p^2}) + o(1)
\end{eqnarray}
at the semi-classical limit ($\lambda \gg 1$ and  $s_\ell \gg 1$) and then reproduces the area law up to a small and real regulator $\varepsilon$. In the right hand side of (\ref{asymp naive}), $o(1)$ refers to a function
of $a_H$ in Planck units. The more striking observation is that we recover immediately the area law with the natural choice $\gamma = \pm i$. This situation strongly contrasts with what happens in the real $\gamma$
case. Nonetheless, this observation raises several questions and remarks.

\begin{enumerate}
\item The method  proposed to perform the analytic continuation needs a finite regulator $\varepsilon$ to be well defined, otherwise it diverges. What it the deep meaning of this regulator? How is it fixed to a non-zero
value? 
\item One analytic continuation was proposed. Is it  uniquely defined?
\item For the last remark, let us assume that the representations $j_\ell$ are all equal and fixed to the value  $j$. In that case, we notice immediately that, at the classical limit, 
the number of microstates is polynomial in $j$ when the representations are discrete whereas it increases exponentially in $j$ when it is complex, $j=-1/2+is$ where $s$ is continuous.  
This remark generalizes immediately to the case where the representations $j_\ell$ are different. A similar phenomenon exists in Chern-Simons theory where the quantum amplitudes
(which  are given by the colored Jones polynomial up to eventual normalizations) behave in total different ways at the semi-classical limit when $k$ is discrete or not. This observation is crucial  for the celebrated volume conjecture \cite{Witten analytic, Kashaev}. 
\end{enumerate}

The last remarks suggests that our question present strong similarities with some important questions related to the quantum amplitudes of Chern-Simons theories.
Therefore, we should look at the analytic continuation of Chern-Simons theory which is essential to understand the volume conjecture according to Witten \cite{Witten analytic}. 
This strategy has been successfully applied for our purpose. As we are going to show in Section \ref{complex analysis}, we have adapted the complex analysis techniques presented in
great detail in \cite{Witten analytic} to define properly the analytic continuation of (\ref{dimCS}). Interestingly, we have recovered in that way exactly the leading order term (\ref{asymp naive}) obtained in
the naive analytic continuation, but the quantum corrections are different. Furthermore,  these methods have enabled us to answer the two first questions above. The regulator $\varepsilon$ can be shown to
be a function of the area $a_H$ at the semi-classical limit. The question of the unicity of the analytic continuation is transposed into a question of finding suitable integration contours in the complex plane.
We have shown that there is indeed an ambiguity in the choice of the integration contour, the same one as the ambiguity illustrated by Witten in \cite{Witten analytic} to define the analytic continuation of the Bessel
function (as originally shown in \cite{Berry}). But the ``simplest" choice (and somehow the most natural one) leads to the expected behavior for the number of microstates. Before explaining these results in detail in Section \ref{complex analysis}, we will continue presenting the complex
black hole focussing now on its thermodynamical physical properties. Understanding these properties is essential in the construction and in the asymptotic analysis of the analytic continuation of the number of 
microstates. 

\subsection{Grand canonical partition function:  semi-classical analysis}
\label{subsection partition}
In  statistical physic, the number of microstates (\ref{dimCS}) and its analytic continuation is the building block for computing  the microcanonical partition function of the black hole.
There is no need of notion of energy nor of temperature to define this partition function because we are counting the number of microstates of a totally isolated system.
As a consequence, the microcanonical ensemble is not the good framework to study thermodynamical properties of the black hole and to understand, for instance, how the semi-classical regime is achieved.
On the contrary, the canonical and grand canonical ensemble do enable us to study the thermodynamical limit provided that we could associate to the quantum black hole a proper notion of energy.
Defining a good notion of black hole energy is a difficult problem in quantum gravity mainly because of  the famous problem of time. However, it was shown in \cite{GP,FGP} that there is a well-defined notion of
energy if one introduces a new length scale $L=a^{-1}$ into the scenario. This scale measures the proper distance of a stationary observer from the black hole horizon, and $a$ is the acceleration needed to keep
its distance to the horizon fixed. It is remarkable to notice that such an observer associates an energy to the black hole which is proportional to the area horizon $a_H$ according to the formula
\begin{eqnarray}\label{FGP energy}
E\{(n_\ell,j_\ell)\} = \frac{a_H}{8\pi L} = \frac{\gamma \ell_p^2}{L} \sum_\ell n_\ell \sqrt{j_\ell(j_\ell+1)}\;.
\end{eqnarray}
We have denoted by $n_\ell$ the number of punctures crossing the horizon colored by $j_\ell$.
This expression is different from the energy (the mass) of the black hole far away from the horizon which is proportional to the square root of $a_H$. Note that this formula holds for $j_\ell$
integers or $j_\ell$ complex with respectively $\gamma$ real and $\gamma=\pm i$. 

Now, we have the main ingredient to compute the canonical partition function $Q_n(\beta)$ defined for a fixed number $n$ of punctures as a function of the inverse temperature $\beta$. 
The equivalence between the microcanonical and the canonical ensembles implies that,
\begin{eqnarray} 
Q_n(\beta)= \sum_{\{n_\ell\}} \frac{1}{\prod_\ell n_\ell !}  \; g\{(n_\ell,j_\ell)\} \exp(-\beta E\{(n_\ell,j_\ell)\})\label{canonical}
\end{eqnarray}
where the sum over ${n_\ell}$ is such that  $\sum_\ell n_\ell = n$. Here, the partition function is obtained as a sum over families $\{n_\ell\}$ which define microstates $\{(j_\ell,n_\ell)\}$ such that the color $j_\ell$
appears $n_\ell$ times.  The degeneracy factor $g\{(n_\ell,j_\ell)\}$, it is given by the dimension of the Chern-Simons Hilbert space (\ref{dimCS})
or by its analytic continuation for the microscopic black hole configuration $\{(n_\ell,j_\ell)\}$.
The presence of the factorials enables us to implement indistinguishability between the $n_\ell$ punctures with the same color $j_\ell$.
The appropriate way to implement the indistinguishability would have been to make a choice of quantum statistics (bosons, fermions or eventually anyons)  for these punctures which  was done
in \cite{Amit}. For simplicity here, we implement the  indistinguishability by introducing a Gibbs factor in the partition function (the factorials) which is well-known to be a good approximation of the quantum statistics
at the (high temperature) thermodynamic limit. The computation of (\ref{canonical}) does not give a simple closed formula when we take the exact Chern-Simons dimension for the degeneracy factors $g\{(n_\ell,j_\ell)\}$.
Fortunately, it is good enough to consider the expression of $g\{(n_\ell,j_\ell)\}$ for large $j_\ell$ (or equivalently large $s_\ell$ in the complex case) which enables us to understand properly the thermodynamical limit.
With this approximation, the area spectrum becomes linear and the degeneracy factor is either polynomial (\ref{discrete asymp}) in the variables $j_\ell$ when they are discrete whereas it is exponential (\ref{asymp naive}) 
in the variables $s_\ell$ when $j_\ell=(-1+is_\ell)/2$ is complex:
\begin{eqnarray}
&&g\{(n_\ell,j_\ell)\} =G_d \prod_\ell (2j_\ell)^{n_\ell} \;\;\; \text{for} \;\; j_\ell \; \text{discrete} \\
&&g\{(n_\ell,j_\ell)\}=G_c \prod_\ell \exp(\pi n_\ell s_\ell)  \;\;\; \text{for} \;\; j_\ell = (-1+is_\ell)/2
\end{eqnarray}
where $G_d$ and $G_c$ are normalization coefficients. We will not take them into account in this Section, and 
for simplicity, we fix them to $G_d=G_c=1$ (even if $G_c$ depends on the regulator $\varepsilon$). 
Indeed, the normalization factors are not relevant for our purposes here, but we will not neglect them in the following Section. We will show that there are responsible for
the quantum corrections to the entropy.

It is straightforward to make sense of  the partition function $Q_n$ when $j_\ell$ are discrete. A simple  calculation leads to the following exact formula (when the area spectrum is linear)
\begin{eqnarray}\label{canonical3}
Q_n(\beta) = \frac{q(\beta)^n}{n!}   \;\;\;\;\text{with} \;\;\;\; q(\beta)=\sum_{j} 2j \exp(-\beta\frac{\gamma \ell_p^2}{L} j)=\frac{2\exp(\beta \frac{\gamma \ell_p^2}{2L})}{(\exp(\beta \frac{\gamma \ell_p^2}{2L}) -1)^2} \;.
\end{eqnarray}
The canonical partition function is a smooth function of the temperature and then admits no singularities (a part from its divergence at very high temperature $\beta \rightarrow 0$). 
Furthermore, it is immediate to see that the grand canonical partition function and all thermodynamical functions behave also correctly for non infinite values of the temperature. As a consequence,
we do not expect any phase transition and nothing special happens when $j_\ell$ are integers. Thus, this case is physically not relevant.

The situation is much more interesting when the degeneracy is holographic and grows exponentially with the area. 
However, in that case,  the explicit calculation of $Q_n$  is more subtle because there is an uncountable set of possible colors for the punctures. The
expression (\ref{canonical3}) is no more relevant and have to be adapted. In fact, it is immediate to see that (\ref{canonical3}) is equivalent to
\begin{eqnarray} 
Q_n(\beta)= \sum_p \sum_{n_1,\cdots,n_p} \frac{1}{\prod_\ell n_\ell !}  \sum_{j_1>\cdots > j_p}\; g\{(n_\ell,j_\ell)\} \exp(-\beta E\{(n_\ell,j_\ell)\})\label{canonical2}
\end{eqnarray}
where the integration measure $\int ds$ is the Lebesgue measure on $[s_0,\infty]$, $s_0>0$ being an eventual area gap.
Such an expression is easily extended to the case where $j_\ell=(-1+is_\ell)/2$:
\begin{eqnarray}
Q_n(\beta)=\sum_p \sum_{n_1,\cdots,n_p} \frac{1}{\prod_\ell n_\ell !}    \int ds_1 \int^{s_1} ds_2 \cdots \int^{s_{p-1}} ds_p \prod_\ell \exp(- x n_\ell s_\ell) 
\end{eqnarray}
with $x=\beta \ell_p^2/2L -\pi$. 
Note that the sum over $(n_1,\cdots,n_p)$ is constrained to $\sum_\ell n_\ell =n$.
To perform the calculation, we start by replacing the integrals by the limit of their Riemann sums according to
\begin{eqnarray}
\int ds f(s) \; = \; \lim_{\epsilon \rightarrow 0} \epsilon \sum_{k} f(k\epsilon)
\end{eqnarray}
 where the sum runs over the integers. In that way, the calculation of the partition function
reduces exactly to the previous one:
\begin{eqnarray}
Q_n(\beta)= \lim_{\epsilon \rightarrow 0} \epsilon^n \sum_{\{n_\ell\}}  \prod_\ell \frac{1}{ n_\ell !}     \exp(- x n_\ell k_\ell \epsilon) = \frac{q(\beta)^n}{n!} \;\;\;
\text{with} \;\;\; q(\beta)=  \frac{\exp(-x s_0)}{ x} 
\end{eqnarray}
where $s_0$ is eventually a non-zero  gap in the continuous area spectrum. As we are going to see, the presence of the area gap does not affect the thermodynamical limit.
Contrary to the discrete case,  the partition function converges only for sufficiently law temperature, more precisely for $\beta > \beta_U$ where $\beta_U=2\pi L/\ell_p^2$ is the 
Unruh temperature for the near horizon observer. The divergence of the partition function at the Unruh temperature is a signature of a phase transition. In fact,
the black hole becomes classical at $\beta_U$ in the sense that its macroscopic area  becomes infinitely large.

In the regime we consider here, we do not expect the number of punctures to be strictly conserved. Hence, it is best to use the grand canonical ensemble. 
Furthermore, we argue in \cite{Amit} that the system of punctures   is very much analogous to a system 
of photons where the photon number is not conserved, hence the chemical potential should vanish.  For that reason we set the fugacity $z = 1$ here (we will consider a non vanishing chemical potential later) 
and the grand canonical partition function
is given by:
\begin{eqnarray}
Z(\beta) = \sum_n Q_n = \exp(q)=\exp(\frac{\exp(-xs_0)}{x}) \;.
\end{eqnarray}
As a consequence, the expression of the mean value $\langle a_H \rangle $ of the horizon area 
\begin{eqnarray}\label{a large}
\langle a_H \rangle = -8\pi L  \frac{\partial}{\partial \beta} \log Z(\beta) =  \frac{4\pi \ell_p^2}{x^2}( 1 +  o(1) )
\end{eqnarray}
enables us to see that the thermodynamical limit is achieved when $x \rightarrow 0$ or equivalently when $\beta$ approaches the inverse Unruh temperature $\beta_U$. In this regime,
the area becomes macroscopic. Even if the number of punctures is not conserved, we can compute their mean  number  $\langle n \rangle$ and look for its asymptotic behavior
at the thermodynamic limit ($x \ll 1$):
\begin{eqnarray}\label{n large}
\langle n \rangle = \frac{1}{Z}\sum_n n Q_n =  \frac{q \exp q }{Z} = \frac{1}{x} (1+o(1)) \sim \sqrt{\frac{\langle a_H \rangle}{4\pi \ell_p^2}} \;. 
\end{eqnarray}
The symbol $\sim$ means equivalent. Therefore, we see that the number of punctures is large at the semi-classical limit and grows as the square root of the macroscopic horizon area.  It can be shown \cite{Amit}  that the average complex 
 $\langle s \rangle$ becomes large as well at the semi-classical limit and grows as $\langle a_H \rangle^{1/2}$ (for the bosonic or Maxwell-Boltzmann statistics).  These results justify to define the thermodynamical
 limit as the regime where the colors and the number of punctures become large.
 
\section{Complex black holes from complex analysis}
\label{complex analysis}
In this Section, we give a rigorous  construction of the analytic continuation of the dimension of Chern-Simons theory Hilbert space (\ref{dimCS}) for a punctured 2-sphere.
The dimension (\ref{dimCS}) is viewed as a function of the level $k$ and  the spin dimensions $d_\ell=2j_\ell +1$, and the analytic continuation refers to taking $k$ and $d_\ell$ 
away from integer values. More precisely, we will be concerned with the case where $k$ and $d_\ell$ are purely imaginary. As we have seen from the relationship between Chern-Simons
theory and black holes in Loop Quantum Gravity, these ``complex" spins case correspond to fixing the Barbero-Immirzi parameter to the value $\gamma=\pm i$ keeping at the same time
the area spectrum (\ref{area spectrum}) real and positive. 

First of all, we express (\ref{dimCS}) as an integral ${\cal I}_k$ of a holomorphic function along a contour in the complex plane. We study the effect of sending the level $k$ to a purely imaginary value
$i\lambda$ where $\lambda \in \mathbb R$. We show in particular that, if we want the integral ${\cal I}_k$ to be non-trivial   when $k$ is purely imaginary then $d_\ell$ must take non integer values as well. 
The requirement that the horizon area is still real and positive imposes that $d_\ell$ is also purely imaginary and can be written as  $d_\ell=is_\ell$. 
This enables us to define the analytic continuation of  ${\cal I}_k$, denoted ${\cal J}_\lambda$, by a suitable choice of contour. Finally, we study the asymptotic of  ${\cal J}_\lambda$ when $\lambda$,  
$n$ and $s_\ell$ are large. We start with the case where there is only one type of puncture, all colored with the
same complex spins. Then, we analyze the general case with many colors. 

\subsection{Analytic continuation}
\begin{center}
\begin{figure}[h]
\includegraphics[scale=1]{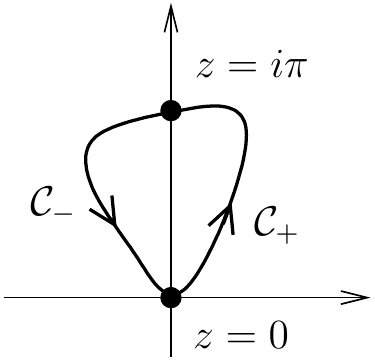}
\caption{The contour $\cal C$ encircles the complex segment $[0,i\pi]$ in the purely imaginary line $i\mathbb R$. 
The points $z=0$ and $z=i\pi$ belong to $\cal C$.
To study the limit when $k \rightarrow \infty$, one decomposes the contour into two parts
${\cal C}_\pm$ defined on the right or left half-planes  ${\Re }(z)>0$ and  ${\Re }(z)<0$.}
\label{contour}
\end{figure}
\end{center}

Any finite sum can be trivially  interpreted as the sum of residues of a suitable function, and therefore it can be expressed as an integral on the complex plane along a contour
which encircles the poles associated to the residues. Hence, the sum (\ref{dimCS}) can be written as the integral\footnote{One would obtain the same value for the integral if one multiplies the integrand by a function $F(z,k)$ 
which admits no poles in the contour $\mathcal C$ and takes the values $F(z_p,k)=1$ at the poles $z_p=i\pi p/(k+2)$. However, such a function could affect the analytic continuation (for complex $k$). Fortunately, this is not really the case
if one assumes that $F(z,k)$ is analytic in both variables and have no poles at all, which is a very natural requirement. More precisely, two different choices of such functions would lead to analytic continuations which would differ
only by a multiplicative constant. This is easily seen from the formula (\ref{def function}) of the analytic continuation.}
\begin{eqnarray}\label{complex integral}
{\cal I}_k = \frac{i}{\pi} \oint_{\cal C} dz \; \sinh^2(z) \lbrace \prod_{\ell=1}^n \frac{\sinh(d_\ell z)}{\sinh z}\rbrace \coth((k+2)z) 
\end{eqnarray}
where the contour $\cal C$ is illustrated in  (\ref{contour}).  For simplicity, we assume that the points $z=0$ and $z=i\pi$ belong to $\cal C$. Here, we have explicitly indicated the $k$ dependence of the integral, and the dependence on the $d_\ell$ is implicit.
When $d_\ell$ are integer, the poles of the integrand are those of the function $\coth((k+2)z)$, hence they are located at the points $z_p=i\pi \frac{p}{k+2}$ where $p$ is an integer which is not a multiple of $k+2$.
Notice that there is an ambiguity in the choice of the contour. Indeed, one can globally translate $\cal C$ in the direction of the imaginary axe with an arbitrary length $i\alpha \in i\mathbb R$ in order to obtain a new 
contour that encircles the segment $[i\alpha,i(\pi+\alpha)]$. This ambiguity is obviously not relevant when $k$ and $d_\ell$ are discrete due to the periodicity of the integrand in (\ref{complex integral}). 
But it has important consequences
 when these labels are away from integer values. We will discuss this aspect later. A very similar  ambiguity exists in the analytic continuation process of the Bessel function (see \cite{Witten analytic} where this ambiguity is well-explained).

At the large $k$ limit ($k$ being an integer), the integral (\ref{complex integral}) simplifies and gives, as expected, the classical volume $\text{Vol}$ of the moduli space of $SU(2)$ flat connections on the punctured
2-sphere (defined from the symplectic structure on the moduli space):
\begin{eqnarray}
\text{Vol}={\cal N}_\infty (d_\ell) = \frac{2}{\pi} \int_0^\pi d\theta \; \sin^2(\theta) \prod_{\ell=1}^n \frac{\sin(d_\ell \theta)}{\sin \theta}.
\end{eqnarray}
This can be seen directly from the sum (\ref{dimCS}) which is in fact the Riemann sum of the previous integral with $\pi/(k+2)$ as elementary interval in $[0,\pi]$. Interestingly, it can also be seen from the integral expression ${\cal I}_k$,
which is more helpful in the perspective of the analytic continuation to complex $k$. Indeed, for any $z \in \mathbb C/i\mathbb R$,  when $k \rightarrow \infty$, the evaluation $\coth((k+2)z)$ tends to a discontinuous function $\chi(z)$:
$\chi(z)=1$ if ${\Re }(z)>0$ and $\chi(z)=-1$ if ${\Re }(z)<0$. A priori, $\chi$ is not defined on $i \mathbb  R$.
Hence, to evaluate the integral (\ref{complex integral}) when $k$ is infinite, one must decompose the contour ${\cal C}$
as a product of two contours ${\cal C}_+$ and   ${\cal C}_-$ defined respectively on the half-planes ${\Re }(z)>0$ and  ${\Re }(z)<0$ as illustrated in (\ref{contour}). Using this decomposition, it is immediate to write the integral
(\ref{complex integral}) as a sum ${\cal I}_k={\cal I}_k^++{\cal I}_k^-$ where the two components are identical at $k=\infty$. As a consequence,
\begin{eqnarray}
{\cal I}_\infty = 2 \, {\cal I}_\infty^+ = \frac{2i}{\pi} \int_{{\cal C}_+} dz  \; \sinh^2(z)  \prod_{\ell=1}^n \frac{\sinh(d_\ell z)}{\sinh z} = \text{Vol} \;.
\end{eqnarray}
The last equality has been obtained by shrinking the contour ${\cal C}_+$ to the interval $[0,i\pi]$. 

\medskip

What happens when the level $k$ becomes complex? The poles of the integrand in (\ref{complex integral}) are no longer located on the imaginary axe but they still belong to a straight-line passing by the origin whose equation is
$z(t)=t i\pi/(k+2)$, $t \in \mathbb R$. As $k$ is assumed to be large, we can approximate $k+2$ by $k$ (in fact, this can be viewed by a redefinition of $k$). 
When $k$ is purely imaginary, i.e. $k=i\lambda$
with a positive (and hence large) $\lambda$, the poles of the integrand are real. As a consequence, the integral (\ref{complex integral}) along the contour ${\cal C}$ represented in the figure (\ref{contour}) vanishes identically, and the analytic continuation of the Chern-Simons dimension to complex levels is a non-sense (at least it is totally useless).   There are two  options to make this analytic continuation relevant. The first one consists in changing the integration contour
and choosing a new contour that encircles a part of the real axe. Unfortunately, there are no natural choice for the new contour and we do not see any mathematical or physical guide to select one contour rather than another. 
The second option is physically and mathematically much more interesting. It is based on the observation that one cannot make $k$ purely imaginary without making the dimensions $d_\ell$ purely imaginary at the same time
if one wants to keep the area spectrum real. Indeed, $k$ purely imaginary means that the Barbero-Immirzi parameter is fixed to $\gamma=\pm i$. 
Therefore, we replace $d_\ell$ by $is_\ell$ where $s_\ell \in \mathbb R^+$ and we observe immediately that  (if $n>2$) the integrand in (\ref{complex integral}) admit new poles 
of order $n$ located at the points $z_p=ip\pi$ where $p$ is a non-vanishing integer.  

The pole located at $z_1=i\pi$ belongs to the contour $\cal C$ introduced to define ${\cal I}_k$ (\ref{complex integral}) and therefore the analytic continuation diverges. However, having considered a contour $\cal C$
that passes by $z_1$ (when $d_\ell$ are integers) was just a matter a choice which was convenient to study the large $k$ limit. We could have moved the contour $\cal C$ such that it passes slightly below the
point $z_1$ or above it provided $\cal C$ do not cross one of the $z_p$.  Thus, considering a contour $\cal C$ which passes exactly at $z_1$ 
or slightly above or slightly below gives the same result for the integral (\ref{complex integral}) when $k$ and $d_\ell$ are discrete. However, when $k=i\lambda$ and $d_\ell=is_\ell$, the situation is completely different.
If $z_1 \in \cal C$, then ${\cal I}_k$ diverges as we said whereas ${\cal I}_k$ vanishes identically if $\cal C$ passes  below $z_1$. The only relevant analytic continuation is obtained taking $\cal C$ which passes slightly 
above $i\pi$. From now on, $\cal C$ denotes this particular contour. As a consequence, the analytic continuation of the integral (\ref{complex integral}) is defined by
\begin{eqnarray}\label{analytic continuation}
{\cal J}_\lambda =  {\cal I}_{i\lambda}\vert_{d_\ell=is_\ell}= \frac{i}{\pi} \oint_{\cal C} dz \; \sinh^2(z) \lbrace \prod_{\ell=1}^n \frac{\sinh(is_\ell z)}{\sinh z}\rbrace \coth((i\lambda+2)z) .
\end{eqnarray}
It can equivalently  be expressed as a residue according to the formula
\begin{eqnarray}\label{residue}
{\cal J}_\lambda = -2 \text{Res}(f;i\pi) \;\;\;\text{with} \;\;\;
f(z)=\sinh^2(z) \lbrace \prod_{\ell=1}^n \frac{\sinh(is_\ell z)}{\sinh z}\rbrace \coth((i\lambda+2)z).
\end{eqnarray} 

It is important to remark that there is in fact another ambiguity in the definition of (\ref{analytic continuation}), more precisely in the choice of the contour $\cal C$. When $k$ and $d_\ell$
are integers, we noticed that one can arbitrarily translate $\cal C$ in the direction of the imaginary axe without changing the value of the integral (\ref{complex integral}). This symmetry
in the choice of the contour does  not hold anymore for the analytic continuation we have just defined (\ref{analytic continuation}). If the translation is too large, the contour $\cal C$ will encircle
a new pole of the integrand in (\ref{analytic continuation}) located at a point $z_p=ip\pi$, $p>1$. The value of the integral would be in that case different from the original one (\ref{analytic continuation}) 
and the semi-classical limit of the analytic continuation would be strongly affected. As we will show in the following subsections, the choice we made above, which is the simplest and somehow the most
natural one, will lead to the correct expression for the black hole entropy. Another choice would not enable to recover the area  law. This justifies a posteriori our original definition of the contour $\cal C$.
However, it would be very interesting to find an independent argument to support our choice.

We finish this Section by a brief study of the limit $\lambda \rightarrow \infty$. Understanding this limit is necessary to obtain the semi-classical behavior of (\ref{analytic continuation}). 
In fact, to obtain the semi-classical behavior, one should study  the limits $s_\ell$ large, $n$ large and $k $ large simultaneously. However, it is simple to see that the large $k$ limit commutes 
with the two others when one writes the integral (\ref{analytic continuation}) as follows:
\begin{eqnarray*}  
{\cal J}_\lambda = \frac{i}{\pi} \oint_{\cal C} dz \; \sinh^2(z) \lbrace \prod_{\ell=1}^n \frac{\sinh(is_\ell z)}{\sinh z}\rbrace \left( -1 + \nu_\lambda(z)\right) \;\;\;\text{where} \;\;\;
\nu_\lambda(z)=\frac{2}{1-\exp(-2(i\lambda +2)z)}.
\end{eqnarray*}
Then, ${\cal J}_\lambda={\cal J}_\infty + {\cal J}_{cor}$  where the correction $ {\cal J}_{cor}$ is the integral which contains the function $\nu_\lambda$. 
To see that $ {\cal J}_{cor}$ is totally negligible, it is sufficient to look at the value of $\nu_\lambda$ at the location of  the  pole $z_1=i\pi$. Indeed, using the residue theorem (\ref{residue})
it is immediate to see that there exists a real positive number $M$ such that $ \vert {\cal J}_{cor} \vert < M \vert \nu_\lambda(i\pi)  {\cal J}_\infty \vert $. Furthermore, a direct calculation shows that $\nu_\lambda (i\pi)$ 
decreases exponentially with $\lambda$ according to $\vert \nu_\lambda(i\pi)\vert \simeq e^{-2\pi \lambda}$. 
For that reason, we can consider $k \rightarrow \infty$ first. 
As a consequence, the large $k$ behavior of the analytic continuation (\ref{analytic continuation}) is dominated by
\begin{eqnarray}\label{def function}
{\cal J}_\infty =  \frac{1}{i\pi} \oint_{\cal C} dz \; \sinh^2(z) \prod_{\ell=1}^n \frac{\sinh(is_\ell z)}{\sinh z}.
\end{eqnarray}
It remains to evaluate the large $n$ and large $s_\ell$ asymptotic of this function. This aspect will be studied in great details in the following two subsections.

\subsection{Asymptotic: one color black hole}
This subsection is devoted to study the semi-classical limit of the function (\ref{def function}). We saw in the previous Section  that the horizon area (\ref{a large}) becomes macroscopic at the
semi-classical limit such that the mean number $\langle n \rangle$ of punctures is large (\ref{n large}) and necessarily the mean color $\langle s \rangle$  becomes large as well. In principle, we should study the two limits simultaneously
to obtain the correct asymptotic behavior of the analytic continuation of the number of microstates. This is quite an involved problem to analyze when there is an arbitrary number of colors $s_\ell$. 
The problem is that if we know the semi-classical behavior of the mean color and the mean number of particles but we do not know how each color individually behaves at this limit. It might be for instance that the 
growth rate of the colors $s_\ell$ with the horizon area depends on the colors themselves. A similar remark holds for the number of punctures $n_\ell$ as well. 
For that reason, we will exclusively consider black hole configurations with a finite number $p$ of colors $(s_1,\cdots,s_p)$. Furthermore, we assume that these colors $s_\ell$ together with the number $n_\ell$ of punctures colored by $s_\ell$  
becomes large at the semi-classical limit in the same way.  The case with an arbitrary number of punctures will not be studied in this article, but it will presumably lead to the same qualitative (and even quantitative) results as the ones
we are going to establish. For clarity reasons, we start with a configuration where the $n$ punctures have the same color $s$.  In that case, the function (\ref{def function}) reduces to the form
\begin{eqnarray}\label{one color}
{\cal J}_s(n)=  \frac{1}{i\pi} \oint_{\cal C} dz \; \sinh^2(z) \left(\frac{\sinh(is z)}{\sinh z}\right)^n
\end{eqnarray}
and the horizon area is given by $a_H=4\pi \ell_p^2 ns$.

\subsubsection{Saddle point approximation}
\label{saddle}
 We assume that $s$ is fixed but large and we study the asymptotic when $n$ goes to infinity. To study the large $n$ limit, it is more convenient to write (\ref{one color}) as the integral
 \begin{eqnarray}\label{one color action}
{\cal J}_s(n)=  \oint_{\cal C} dz \; \mu(z)  \exp(n{\cal S}(z)) \;\;\;\; \text{where} \;\;\; {\cal S}(z)=\log\left( \frac{\sinh(is z)}{\sinh z}\right) 
 \end{eqnarray}
and $n$ plays the role of the  inverse temperature. The measure factor $\mu(z)=\sinh^2(z)/i\pi$ will play a central role in the calculations of  the quantum corrections of the entropy.
The study of the asymptotic relies on the analysis of the critical points $z_c$ of the action ${\cal S}(z)$. They are given by the solutions of the equation
\begin{eqnarray}\label{tan equation}
\tan(sz)= s \tanh(z) \;.
\end{eqnarray}
The precise analysis of these critical points have been done in the appendix. In particular, it is shown that critical points are either on the real line $\mathbb R$ or on the purely imaginary axe
$i\mathbb R$. All these points could contribute to the asymptotic of the function (\ref{one color}). However, the dominant contribution comes from the critical point $z_c$ located at the vicinity of the pole $z_1=i\pi$.
A short analysis shows that $z_c=i(\pi + \epsilon)$ where $\epsilon = s^{-1}(1+ o(s^{-1}))$. As $s$ is supposed to be large, we will approximate $\epsilon$ by $s^{-1}$. The corrections $o(s^{-1})$ we are neglecting will
not have any relevance in the study of the asymptotic neither at the level of the leading order nor at the level of the quantum sub-leading corrections (for the entropy).  

Hence, we can make use of the gaussian approximation at the vicinity of $z_c$ to obtain the asymptotic of (\ref{one color}).  For this, we need to evaluate the function ${\cal S}$ at the critical points 
\begin{eqnarray}\label{Sc}
{\cal S}(z_c)=\log \left( -i \frac{\sinh(s\pi +1)}{\sin(1/s)}\right) \simeq  -i\frac{\pi}{2} + s \pi + \log(\frac{se}{2}) 
\end{eqnarray}
but also its second derivative ${\cal S}''$ and the measure $\mu$
\begin{eqnarray}\label{deriv Sc}
{\cal S}''(z_c)= -(s^2 + 1) + \frac{1}{\tanh^2(z_c)} - \frac{s^2}{\tan^2(sz_c)} \simeq -s^2 \;\;,\;\;\;\;\;
\mu(z_c)=\frac{1}{i\pi}\sinh^2(z_c) \simeq \frac{i}{\pi s^2} .
\end{eqnarray}
We used the principal branch for the complex logarithm in (\ref{Sc}) and the equation (\ref{tan equation}) to simplify the expression of ${\cal S}''(z_c)$ in (\ref{deriv Sc}).
In these formulae, the symbol $\simeq$ means that we keep the relevant dominant contributions when $s$ is large. The neglected terms do not contribute at all to the asymptotic. 
As a consequence, according to the saddle point approximation, when $n$ is large, the complex integral is equivalent  to the following gaussian integral
\begin{eqnarray}
{\cal J}_s(n) \sim {\cal G}_s(n)= \mu(z_c) \exp(n\, {\cal S}(z_c)) \int dx \, \exp(n \, {\cal S}''(z_c)\frac{x^2}{2}) 
\end{eqnarray}
which can be explicitly evaluated to obtain the asymptotic expression for ${\cal J}_s(n)$
\begin{eqnarray}\label{asymp gauss}
{\cal G}_s(n)= \sqrt{\frac{2}{\pi}} \frac{1}{s^3 \sqrt{n}}\left(\frac{se}{2}\right)^n \exp{ (\pi ns + i(1-n) \frac{\pi}{2})}.
\end{eqnarray}
As the horizon area is given by $a_H=4\pi \ell_p^2ns$ in this case, we see immediately that the leading order term of the microcanonical entropy $S=\log{\cal G}_s(n)$ reproduces 
exactly the area law $S \sim a_H/4\ell_p^2 $ when $a_H$ is large with the proper $1/4$ factor.  Before going further, let us make some important remarks concerning this result.
\begin{enumerate}
\item The presence of the phase $\exp(i(1-n)\frac{\pi}{2})$ in the asymptotic expansion raises questions. Indeed, we expect the analytic continuation to be a non-negative real function for it to define a number
of black hole microstates. We propose two options to resolve this puzzle. First, it might be that only the modulus $\vert  {\cal J}_s(n) \vert$ could be interpreted as the number of microstates and then there is no more problem with the phase. 
Another possibility is that the number of punctures $n$ is not totally arbitrary and one must choose it such that the phase is fixed to one, which implies that $1-n=4m$ where $m$ is a natural number. The interpretation of such a condition
is not clear but one of its interesting consequence is that the minimal number of punctures is $n=3$. For the time being, we will chose the first option and then we will simply omit the phase in the asymptotic. 
\item Up to a global multiplicative factor, the asymptotic obtained rigorously here has an expression very similar to the one obtained ``intuitively" or ``naively" in \cite{FGNP} and briefly recalled in the introduction (\ref{asymp naive}). 
In particular, we can clearly identify the regular $\varepsilon$ introduced in \cite{FGNP} to avoid having divergences in the analytic continuation with the term $es/2$ which appears at the power $n$ in the formula
above. Thus, the regulator 
is not arbitrary and its relevance receives here a clear justification. This gives an answer to the first question raised in subsection (\ref{gamma = i}).  
\item So far we have a clear interpretation of the leading order term $S \sim a_H/4\ell_p^2$ in the large $a_H$ expansion of the entropy $S$ but the status of the subleading terms is still enigmatic at this point.  If we remind the analysis of
\cite{Amit} recalled in subsection (\ref{subsection partition}), both the average number of punctures and the average color grow like $\sqrt{a_H}$. Thus, it is expected that, at the semi-classical limit, we can replace $n$ by $n=\nu \sqrt{a_H}/\ell_p$
and the color $s$ by $s=\sigma \sqrt{a_H}/\ell_p$ where $\nu$ and $\sigma$ are numerical constant. If this is the case, the analytic continuation of entropy behaves at the classical limit as 
\begin{eqnarray}\label{sigma}
S(a_H) \sim \frac{a_H}{4\ell_p^2} + \frac{\nu}{2} \frac{\sqrt{a_H}}{\ell_p}\log(\frac{a_H}{\ell_p^2}) + \nu\log(\frac{\sigma e}{2}) \frac{\sqrt{a_H}}{\ell_p} -\frac{7}{4} \log(\frac{a_H}{\ell_p^2}) + {\cal O}(1).
\end{eqnarray}
where the phase in (\ref{asymp gauss}) has been disregarded. Due to the presence of the second term in the previous expansion, quantum corrections seem a priori larger than the ones found in \cite{Amit}.
The reason of these discrepancies is simple and relies on the fact that we have not implemented the indistinguishability so far. An immediate way to implement the indistinguishability consists in adding a Gibbs factor
in the definition of the number of microstates. Concretely, we replace ${\cal J}_s(n)$ by ${\cal J}_s(n)/n!$. Finally, the Stirling formula leads to the following expansion of the microcanonical entropy
\begin{eqnarray}\label{micro entropy}
S_{micro}(a_H) \sim \frac{a_H}{4\ell_p^2} + \nu \log(\frac{e^2\sigma}{2\nu}) \frac{\sqrt{a_H}}{\ell_p} - 2 \log(\frac{a_H}{\ell_p^2})
\end{eqnarray}
which agrees with \cite{Amit} as expected.  We also obtain logarithmic corrections. Nonetheless, at this point  we cannot trust completely the quantum corrections and particularly the numerical coefficients in each subleading terms
in the large $a_H$ expansion of the entropy. Indeed, we have assumed here that $n$ and $s$ are exactly proportional to $\sqrt{a_H}$ with no subleading corrections. But, as we are going to see, subleading corrections
do exist and slightly modify the quantum corrections of the entropy. 
\end{enumerate}
Up to now, the analysis is microcanonical and the asymptotic behavior of $n$ and $s$ is essentially based on an analogy with \cite{Amit}. Hence, the parameter $\nu$ and $\sigma$ are unfixed, and, as we have just said,
the precise form of $n$ and $s$ in terms of $a_H$ is unknown. 
To get a more accurate expression of $n$ and $s$ and  explicit values of these two parameters $\nu$ and $\sigma$, 
it is necessary to perform a grand canonical analysis of the complex black hole. This is indeed the good framework to compute the mean values
of the number of punctures, the mean color and the quantum corrections to the entropy. 

\subsubsection{Partition function}
Here, we consider the black hole as a gas of indistinguishable punctures. 
As proposed in \cite{Amit}, we first assume that the chemical potential of these punctures vanishes. As we are going to see later, considering a non-vanishing chemical potential
will appear very interesting concerning the quantum corrections of the entropy.
For the time being, the grand canonical partition function depends only on the (inverse) temperature $\beta$ and is defined by
\begin{eqnarray}\label{fonction Zs}
Z(\beta)= \int ds \; Z_s(\beta) \;\;\; \text{with} \;\;\; Z_s(\beta) =\sum_n \frac{g(n,s)}{n!} \exp(-\beta E(n,s))
\end{eqnarray}
where the energy (\ref{FGP energy}) is $E(n,s)=\frac{\ell_p^2}{2L}ns$ and the degeneracy factor $g(n,s)$ should be given by the integral ${\cal J}_s(n)$ (\ref{one color}). 
In the series $n$ runs over non-zero integers and the integral over $s$ is the Lebesgue measure on $[s_0,\infty[$ where $s_0$ is an eventual non-zero area gap.
There is no explicit and simple expression for $Z(\beta)$ when $g(n,s)$ is exactly given by (\ref{one color}): the sum over $n$ can be performed but the integral over $s$ 
is difficult to handle.  For simplicity, we will approximate $g(n,s)$ by the large $n$ asymptotic of ${\cal J}_s(n)$, up to the phase factor and the irrelevant $\sqrt{2/\pi}$ multiplicative factor which will be omitted.
Hence, the degeneracy is
\begin{eqnarray}
g(n,s) =  \frac{1}{s^3 \sqrt{n}}\left(\frac{se}{2}\right)^n \exp{ (\pi ns )}.
\end{eqnarray}
The partition function defined with this degeneracy is in fact equivalent to the exact partition function at the thermodynamical limit. We will come back to this statement later. To simplify the 
expression of the partition function, we start with the calculation of $Z_s(\beta)$:
\begin{eqnarray}\label{variable q}
Z_s(\beta) = \frac{1}{s^3}\sum_{n=1}^\infty \frac{1}{\sqrt{n}} \frac{q^n}{n!} \;\;\;\; \text{with} \;\;\; q=\frac{se}{2} e^{-xs} \;\;\text{and} \;\; x=\beta \frac{\ell_p^2}{2L} -\pi \;.
\end{eqnarray}
The thermodynamical limit is defined by $x \rightarrow 0$ which is equivalent to taking $\beta \rightarrow \beta_U$ where $\beta_U$ is the inverse
Unruh temperature. It is easy to see that the grand canonical partition function is defined only for $x>0$. Indeed, when $x$ is non positive, the integral over $s$
diverges. This is the reason why the exact partition function is equivalent to the one where the degeneracy is given by $g(n,s)$ as $x$ approaches $0$.

Now, we are going to see that the mean value of the area scales as $x^{-2}$ at the vicinity of $x=0$ whereas we argued that, at the same time,
the mean  color $s$ increases with $\sqrt{a_H}$, and then it scales as $x$. As a consequence, the variable $q$ defined above (\ref{variable q}) becomes
large at the semi-classical limit. This motivates the study of the large $q$ expansion of the series in (\ref{variable q}). To do so, we first express the series as
follows:
\begin{eqnarray}
I(q) \equiv  \sum_{n=1}^\infty \frac{1}{\sqrt{n}} \frac{q^n}{n!} = \frac{1}{\sqrt{\pi}} \sum_{n=1}^\infty  \int du \, \frac{(q \exp(-u^2))^n}{n!}= \frac{1}{\sqrt{\pi}}  \int du \left( \exp(qe^{-u^2}) -1 \right) \nonumber.
\end{eqnarray}
The large $q$ limit of this last integral is governed by the critical points of the function $f(u)=e^{-u^2}$. There is only one critical point (at $u=0$) and therefore we obtain the following large $q$ expansion
\begin{eqnarray}
I(q) = I_\infty(q)(1+o(1)) \;\;\; \text{with} \;\;\; I_\infty(q) =  \frac{1}{\sqrt{\pi}}  \int du \exp(q(1-u^2)) = \frac{\exp(q)}{\sqrt{q}} .
\end{eqnarray}
As a consequence, the expression of $Z_s(\beta)$ reduces to the form
\begin{eqnarray}
Z_s(\beta) = \frac{1}{s^3} \frac{\exp(q)}{\sqrt{q}} (1+o(1))
\end{eqnarray}
which is particularly interesting to study the semi-classical limit of the partition function. It is indeed now easy to get an equivalent of $Z(\beta)$ when $x$ approaches zero.
First we notice that part which contains $o(1)$   does not contribute  to $Z(\beta)$ at the leading order and then we obtain immediately
\begin{eqnarray}
Z(\beta) \sim Z_0(\beta)=\int \frac{ds}{s^3}  \frac{\exp(q)}{\sqrt{q}} = \int \frac{ds}{s^3} \left( \frac{2}{es} e^{xs}\right)^{1/2} \exp\left( \frac{es}{2} e^{-xs}\right)  .
\end{eqnarray}
Then, we change of variable $s \mapsto u=e^{-xs}$ in the integral 
\begin{eqnarray}\label{integral Z0}
Z_0(\beta) = \sqrt{\frac{2}{e}} x^{5/2} \int \frac{du}{u^{3/2} (-\log u)^{7/2}} \exp(-\frac{e}{2x} u \log u)
\end{eqnarray}
where the integral runs from $0$ to $1-\epsilon$, $\epsilon$ being related to the area gap $s_0$ by $\epsilon=1-e^{-xs_0}$.
 The leading order is obtained by the saddle point approximation at the critical point $u=e^{-1}$ of the exponent:
\begin{eqnarray}\label{equiv Z0}
Z_0(\beta) \sim \sqrt{\frac{2}{e}} x^{5/2} \int du \, e^{3/2} \exp\left( -\frac{e}{2x}(-\frac{1}{e} + e\frac{u^2}{2} )  \right) = \sqrt{8\pi} x^3 \exp(\frac{1}{2x}).
\end{eqnarray}
As a conclusion, the grand canonical partition function of the (one color) black hole behaves as
\begin{eqnarray}
Z(\beta) \sim Z_{sc}(\beta) = \sqrt{\frac{\pi}{8}} \frac{\ell_p^6}{L^3} (\beta-\beta_U)^3 \exp\left( \frac{L}{\ell_p^2(\beta -\beta_U)}\right) 
\end{eqnarray}
at the vicinity of the inverse Unruh temperature $\beta_U$. As expected, it diverges at $\beta_U$ which is the signature to a phase transition from the quantum to the classical regime
of the black hole. The singular point of $Z(\beta)$ at $\beta=\beta_U$ is an essential singularity.

\subsubsection{Thermodynamical limit}
Once we have computed the partition function, we can determine the mean horizon area, the mean number of punctures and the mean color. We are particularly interested in finding the scaling
of this mean values with $x \propto (\beta-\beta_U)$ at the vicinity of the Unruh temperature. 

Concerning the mean horizon area, it is immediately obtained from derivatives of $Z(\beta)$:
\begin{eqnarray}
\langle a_H \rangle =  -8\pi L \, \partial_\beta \log Z(\beta) \sim \frac{2\pi \ell_p^2}{x^2} = \frac{8\pi L^2}{\ell_p^2(\beta - \beta_U)^2}.
\end{eqnarray}
As expected $\langle a_H \rangle$ scales as $x^{-2}$ and becomes macroscopic when $\beta$ approaches $\beta_U$.  

It remains to compute the mean color and the mean number of particles. The latter is the simplest to evaluate. By definition, it is given by
\begin{eqnarray}
\langle s \rangle = \frac{1}{Z(\beta)} \int ds \; s \; Z_s(\beta).
\end{eqnarray}
Using results of the previous subsection, it is straightforward to see that the mean color behaves according to
\begin{eqnarray}
\langle s \rangle \sim \frac{1}{Z_0(\beta)} \sqrt{\frac{2}{e}} x^{3/2} \int^1 \frac{du}{u^{3/2} (-\log u)^{9/2}} \exp(-\frac{e}{2x} u \log u)
\end{eqnarray}
at the vicinity of the Unruh temperature. The same techniques (saddle point approximation) as the one used to simplify (\ref{integral Z0})
 to the expression (\ref{equiv Z0}) enables us to find the behavior of the mean color in the semi-classical regime
 \begin{eqnarray}
 \langle s \rangle \sim \frac{1}{x} = \frac{2L}{\ell_p^2(\beta - \beta_U)} .
 \end{eqnarray}
As expected, the mean color increases when $\beta$ approaches $\beta_U$ and scales as $\sigma \sqrt{\langle a_H \rangle}/\ell_p$. 
Furthermore, now we can extract the value of the coefficient $\sigma$ introduced previously in (\ref{sigma}): $\sigma=(2\pi)^{-1/2}$.

The mean number of punctures is defined by
\begin{eqnarray}
\langle n \rangle = \frac{1}{Z(\beta)} \int ds \; N_s(\beta) \;\;\; \text{with} \;\;\; N_s(\beta) =\sum_n n \frac{g(n,s)}{n!} \exp(-\beta E(n,s))
\end{eqnarray}
where the series $N_s(\beta)$ differs from the function $Z_s(\beta)$ (\ref{fonction Zs}) by the presence of an extra factor $n$.
The calculation of $N_s(\beta)$ and the evaluation of its asymptotic at the vicinity of the Unruh temperature $\beta_U$ is similar
to the calculation of $Z_s(\beta)$.  Indeed, following what we have done for the partition function, we first compute the first term in the large $q$  expansion of $N_s(\beta)$
\begin{eqnarray}
N_s(\beta) =  \frac{1}{s^3} {\exp(q)}{\sqrt{q}} (1+o(1)) ,
\end{eqnarray}
where $q$ has been defined in (\ref{variable q}).  This implies the following equivalences at the vicinity of the Unruh temperature:
\begin{eqnarray}
N_0(\beta) \equiv  \int ds \; N_s(\beta)  \sim \int \frac{ds}{s^3} {\exp(q)}{\sqrt{q}} \sim \sqrt{2{\pi}} x^2 \exp(\frac{1}{2x}).
\end{eqnarray}
We used the same saddle point approximation as in (\ref{equiv Z0}). As a consequence, the mean number of punctures is given by
\begin{eqnarray}
\langle n\rangle \sim \frac{N_0(\beta)}{Z_0(\beta)} \sim \frac{1}{2x} = \frac{L}{\ell_p^2(\beta - \beta_U)}
\end{eqnarray}
at the vicinity of the Unruh temperature. As expected, we see that the mean number of punctures increases with the horizon area according to the law
\begin{eqnarray}
\langle n \rangle \sim \nu \sqrt{\langle a_H\rangle} \;\;\text{with} \;\;\; \nu= \frac{\sigma}{2}={{\frac{1}{\sqrt{8\pi}}}}.
\end{eqnarray}
This determines the second parameter $\nu$ introduced in (\ref{sigma}). As a conclusion of this subsection, we proved that the mean number of punctures and
the mean color scales as $\sqrt{\langle a_H\rangle}$ at the semi-classical limit. This is self-consistent with the saddle point approximation used to find
the asymptotic the number of microstates in subsection (\ref{saddle}).

\subsubsection{Microcanonical vs. Grand canonical entropy: quantum corrections}
Fixing of the parameters $\nu$ and $\sigma$ enables to determine with no ambiguities the first three terms in the large $a_H$ expansion of the microcanonical entropy with
the Gibbs factor (\ref{micro entropy}). To do so, we recall that the gas of punctures  is described at the semi-classical limit by the partition function
\begin{eqnarray}\label{Zsc}
Z_{sc}(\beta) = \sqrt{8\pi} x^3 \exp(\frac{1}{2x}) .
\end{eqnarray}
For simplicity, we will omit the brackets to denote the mean quantities as the area, the number of punctures and the color.
From this expression, we easily exhibit the mean horizon area $a_H$ as a function of $x$ 
\begin{eqnarray}
a_H = 2\pi \ell_p^2 (\frac{1}{x^2} - \frac{6}{x})
\end{eqnarray}
where now we do not neglect consider only the leading order term, but we take into account the subleading corrections. These corrections will be important to compute the large $a_H$
expansion of the microcanonical entropy. Expressing $x$ in terms of $a_H$ allows to determine $n$ and $s$ as functions of $a_H$ according to
\begin{eqnarray}\label{affine s and n}
s = 2n = \sqrt{\frac{a_H}{2\pi \ell_p^2}} + 3 + o(1).
\end{eqnarray}
Now, we are ready to compute the microcanonial entropy replacing these previous expressions in the formula (\ref{asymp gauss}):
\begin{eqnarray}
S_{micro}(a_H) & = & 2\pi n^2 + 2n - 4 \log n + o(\log(n)) \nonumber \\
& = &  \frac{a_H}{4\ell_p^2}+ K \sqrt{\frac{a_H}{\ell_p^2}} - 2 \log(\frac{a_H}{\ell_p^2}) + o(\log(\frac{a_H}{\ell_p^2}))
\end{eqnarray}
where $K=(1+{\color{red}{6}}\pi)/\sqrt{2\pi}$.
As in \cite{Amit}, the subleasing corrections to the area law starts with a term proportional to $\sqrt{a_H}$. In addition, we have here logarithmic corrections.
Note the discrepancies between this expression for the entropy and the one we would have obtained from (\ref{micro entropy}) replacing $\nu$ and $\sigma$
by their expressions found above. The reason comes from the fact that now we have not neglected to subleading terms in the large $a_H$ expansion of the mean
number of punctures $n$ and the mean color $s$ (\ref{affine s and n}).

It is interesting to compare this expression to the grand canonical entropy. As we have argued, we start with a vanishing chemical potential for the punctures.
Hence, the grand canonical entropy simply depends on the temperature and is defined by
\begin{eqnarray}
S_{grand}= \beta U + \log {Z} = \beta \frac{\langle a_H \rangle}{8\pi L} + \log Z.
\end{eqnarray}
There is no simple formula for the grand canonical entropy in general. However, close to the Unruh temperature, we can replace $Z$ by
its semi-classical expression (\ref{Zsc}) and a direct calculation leads to 
\begin{eqnarray}
S_{grand} & = & 2\pi n^2 + 2(1-3\pi)n + 3\log(n) + o(\log(n)) \nonumber \\
  & = & \frac{a_H}{4\ell_p^2} + \sqrt{\frac{a_H}{2\pi \ell_p^2}} - \frac{3}{2} \log(\frac{a_H}{\ell_p^2}) + o(\log(\frac{a_H}{\ell_p^2})) 
\end{eqnarray}
where we used the relations $2n=1/x$ and $\beta=\frac{2L}{\ell_p^2}(x+\pi)$.
Of course, the leading order terms of the two entropies agree (which is a consequence of the equivalence
of the statistical ensembles) but the subleading terms differ essentially due to the fluctuations of the area in the grand canonical ensemble.

\subsubsection{Chemical potential and logarithmic corrections}
In general, it is expected that quantum corrections to the entropy are logarithmic. Here, they are much larger because they start with a $\sqrt{a_H}$ term and then comes the logarithmic
term. Let us note that $\sqrt{a_H}$ is proportional to the mean number of punctures  $n$  which means that considering a non trivial chemical potential $\mu$ for the gas 
of punctures might cancel the $\sqrt{a_H}$ term of the entropy and leaves us with logarithmic corrections only. We are going to show that, for a particular choice of $\mu$,
this is indeed the case. Note that a statistical analysis of black holes in Loop Quantum Gravity with a chemical potential was done in \cite{Vaz} in the case of a real Barbero-Immirzi
parameter $\gamma$.

The grand canonical  partition function in the presence of a chemical potential $\mu$ is very similar to (\ref{fonction Zs})
and can be defined as a function of $\beta$ and the fugacity $z=\exp(\beta \mu)$
\begin{eqnarray}
Z(\beta,z) =  \int \frac{ds}{s^3} \sum_{n=1}^\infty \frac{1}{\sqrt{n}} \frac{Q^n}{n!} \;\;\;\; \text{with} \;\;\; Q=\frac{se}{2} z e^{-xs} 
\end{eqnarray}
Note that $z$ enters in the definition of the variable $Q$ (denoted $q$ in the absence of chemical potential). 
The analysis of its behavior close to the Unruh temperature is now immediate and leads to the following asymptotic expansion:
\begin{eqnarray}
Z(\beta,z) \sim Z_{sc}(\beta,z)=\sqrt{8\pi} \frac{x^3}{{z}} \exp(\frac{z}{2x}) .
\end{eqnarray}
As previously, we will replace the exact partition function by its semi-classical approximation.
The mean area $a_H$ and the mean number of punctures $ n $ at the vicinity of $\beta_U$ follow immediately
\begin{eqnarray}
n  = {z}\partial_z \log Z_{sc}(\beta,z)  = \frac{z}{2x} - 1  \;\;\;\; \text{and} \;\;\;\;
 a_H  =-8\pi L \partial_\beta \log Z_{sc}(\beta,z) = 4{\pi}\ell_p^2 (\frac{z}{2x^2 } - \frac{3}{x}). 
\end{eqnarray}
Notice that we recover what has been computed in previous subsections when $z=1$. We have now all the ingredients to compute the large $a_H$ expansion of the grand canonical entropy
\begin{eqnarray}
S_{grand} = \beta \frac{a_H }{8\pi L} + \log Z(\beta,z) -\beta \mu n \, .
\end{eqnarray} 
We start by expressing the entropy in term of the variable $x$. Using the approximations
\begin{eqnarray}
z=z_U(1+\frac{\mu \beta_U}{\pi}x) \,\;\;\;\; \text{and} \;\;\; \beta=\beta_U(1+ \frac{x}{\pi})
\end{eqnarray} 
where $z_U=\exp(\mu \beta_U)$, we obtain
\begin{eqnarray} 
S_{grand}= \frac{\pi z_U}{2x^2} + (\frac{z_U}{\pi} -3) \frac{\pi}{x} + 3\log (x) + o(\log(x))
\end{eqnarray}
where $o(\log(x))$ is in fact a constant . When we replace $x$ by its expression in terms of $a_H$, we get the following large $a_H$ expansion
\begin{eqnarray}
S_{grand}=  \frac{a_H}{4\ell_p^2} - \frac{3}{2} \log(\frac{a_H}{\ell_p^2}) + \frac{z_U}{2x}(2-\mu \beta_U) + o(\log(\frac{a_H}{\ell_p^2})).
\end{eqnarray}
As a consequence, when the chemical potential is fixed at $\mu=2T_U$ where $T_U$ is the Unruh temperature, the grand canonical entropy has only logarithmic corrections
\begin{eqnarray}\label{grand entropy}
S_{grand}(a_H) = \frac{a_H}{4\ell_p^2} - \frac{3}{2} \log(\frac{a_H}{\ell_p^2}) + o(\log(\frac{a_H}{\ell_p^2}) )
\end{eqnarray}
when the area is large. Note that the logarithmic corrections comes with the numerical factor $-3/2$ which is identical to the one obtained for  black hole entropy with a real
Barbero-Immirzi parameter $\gamma$. This observation is particularly interesting because logarithmic corrections are supposed to be universal and independent of $\gamma$
when it is real. The universality of the quantum corrections to the entropy  seems to hold when $\gamma=\pm i$.  This universality is nonetheless limited to the one color black hole model.
When many colors are allowed, the rate of logarithmic corrections change and depends, as we will see, in the number of colors. 

\subsection{Many colors black hole}
In this Section, we are going to show that previous results become widespread to black holes with a finite number $p$ of colors.
Among the $n$ punctures which cross the horizon, $n_1$ are colored with the color $s_1$, $n_2$ with the color $s_2$
and so on.  Therefore, we get the relations:
\begin{eqnarray}\label{total number}
n=\sum_{\ell=1}^p n_\ell \;\;\;\;\text{and} \;\;\;\; a_H=4\pi \ell_p^2 \sum_{\ell=1}^p n_\ell s_\ell .
\end{eqnarray}

\subsubsection{Number of microstates and semi-classical limit}
Exactly as in the one color black hole model, we assume a priori that, at the classical limit, each colors $s_\ell$ and each number of punctures $n_\ell$ are large
for $\ell \in [1,p]$ and check that this assumption is self-consistent. Following the previous strategy for the study of the thermodynamical properties of the system
of punctures, we fix the colors $s_\ell$ to large values and we consider the limit where the numbers $n_\ell$ become large. We suppose that all these numbers
increase at the same velocity, i.e. we can write $n_\ell=\kappa  \nu_\ell$ where $\nu_\ell$ are fixed and $\kappa$ becomes larger and larger at
the semi-classical limit. We will see that necessarily $\kappa$ scales as $\sqrt{a_H}/\ell_p$ where $a_H$ is the (mean) area of the horizon.
In that case, the analytic continuation of the  number of black hole microstates can be written as
\begin{eqnarray}\label{multi color integral}
{\cal J}_{s_\ell}(n_\ell)=  \oint_{\cal C} dz \; \mu(z)  \exp(\kappa{\cal S}_p(z)) \;\;\;\; \text{where} \;\;\; {\cal S}_p(z)=\sum_{\ell=1}^p \nu_\ell \log\left( \frac{\sinh(is_\ell z)}{\sinh z}\right) 
\end{eqnarray}
where the measure $\mu(z)$ and the contour $\cal C$ are the same as in the prior model (\ref{one color action}).

The analysis of the large $\kappa$ asymptotic is very analogous to what we have already done even if it is a bit more involved. For this reason, we will not detail the calculations here.
The study of the asymptotic relies on the analysis of the critical points of ${\cal S}_p$. As it is shown in the appendix, the ``dominant" critical point $z_c$ is located on the imaginary axe.
When the colors are large (what we are assuming from the begining), $z_c$ is very close to the pole $z_1=i\pi$ and can be written as
 \begin{eqnarray}\label{arithmetic mean}
 z_c=i(\pi + \epsilon) \;\;\;\; \text{with} \;\;\;\; \epsilon=\frac{1}{s} +o(s^{-1}) \;\;\; \text{and} \;\;\; s=\frac{\sum_\ell \nu_\ell s_\ell}{\sum_\ell \nu_\ell}.
 \end{eqnarray}
 The expression of $z_c$ is similar to the one color model where the color $s$ is the arithmetic mean color. From now, we will neglect the term $o(s^{-1})$ in the expression of $\epsilon$
 because it will not play any role in the asymptotic of (\ref{multi color integral}). The large $\kappa$ behavior can be captured from  the saddle point approximation which indeed leads to
 the following semi-classical equivalent for the integral (\ref{multi color integral}):
 \begin{eqnarray} 
{\cal J}_{s_\ell}(n_\ell) \sim {\cal G}_{s_\ell}(n_\ell) = \mu(z_c) \exp(n\, {\cal S}_p(z_c)) \int dx \, \exp(n \, {\cal S}_p''(z_c)\frac{x^2}{2}) .
\end{eqnarray}
The evaluation of  $\mu$,  ${\cal S}_p$ and  ${\cal S}_p''$ at $z_c$ is immediate and gives formally exactly the same expression as in the one color model
\begin{eqnarray}
{\cal G}_{s_\ell}(n_\ell) = \sqrt{\frac{2}{\pi}} \frac{1}{s^3 \sqrt{n}}\left(\frac{se}{2}\right)^n \exp{ (\pi ns + i(1-n) \frac{\pi}{2})}
\end{eqnarray}
where $s$ (\ref{arithmetic mean}) and $n$ (\ref{total number}) have already been defined above. 
At the semi-classical limit,  the (analytic continuation of the) number of microstates depends only on the two variables $s$ and $n$
and not on the details of the punctures colors.
As a result, the leading order term of the microcanonical entropy
$S=\log {\cal G}_{s_\ell}(n_\ell) $ reproduces the area law as in the one color model. Although, all the remarks we raised at the end of subsection (\ref{saddle})   
still hold. In particular, to compute the quantum corrections of the entropy and to have a deeper
physical interpretation of the model, it is necessary to study it in the framework of the grand canonical ensemble where we can naturally implement
the indistinguishability. 

\subsubsection{Grand canonical partition function: indistinguishability and chemical potential}
We start directly with the calculation of the grand canonical partition function.  For simplicity, we assume that 
the chemical potential $\mu$ is the same for all the punctures, i.e. $\mu$ does not depend on the color. If we use the usual notation $z=\exp(\beta \mu)$ for the fugacity,
then the grand canonical partition function is given by
\begin{eqnarray}\label{partition multi}
Z_p(\beta,z) = \int d\mu(s_1,\cdots,s_p) \sum_{n_1,\cdots,n_p} \frac{1}{\prod_{\ell=1}^p n_p!} g_p(n_\ell,s_\ell) z^n\exp(-\beta E_p(n_\ell,s_\ell))
\end{eqnarray}
where the degeneracy $g(n_\ell,s_\ell)$ and the black hole energy $E(n_\ell,s_\ell)$ are
\begin{eqnarray}
g_p(n_\ell,s_\ell)= g(n,s)= \frac{1}{s^3 \sqrt{n}}\left(\frac{se}{2}\right)^n \exp{(\pi ns)} \;\;\;,\;\;\; E_p(n_\ell,s_\ell)=\frac{\ell_p^2}{2L} ns \;.
\end{eqnarray}
These two quantities have exactly the same expression as in the one color model. The novelties in the multi-colors model rely on the form of the Gibbs
factor and on the expression of the measure associated to the colors which  is explicitly given by
\begin{eqnarray}\label{measure on s}
\int d\mu(s_1,\cdots,s_p)  = \int_{s_1>s_2>\cdots>s_p} \prod_\ell ds_\ell= \int ds_1 \int^{s_1} ds_2 \cdots  \int^{s_{p-1}} ds_p \,
\end{eqnarray}
where $\int ds$ is the standard Lebesgue measure. The presence of the Gibbs factor and the definition of this integral enable us to implement the indistinguishability of the punctures, but they make
the calculation of $Z_p(\beta,z)$ (even at the vicinity of the Unruh temperature) more involved, as we are going to see.  

To write the partition function is a simpler form, we first replace the measure
(\ref{measure on s}) by the standard multiple Lebesgue measure
\begin{eqnarray}
\int d\mu(s_1,\cdots,s_p)  = \frac{1}{p!}  \prod_{\ell=1}^p \int ds_\ell 
\end{eqnarray}
then we perform the change of variables
\begin{eqnarray}
n t_1=n_1s_1,\;\;\; nt_2=n_1s_1 + n_2s_2,\; \cdots \;\;\; nt_p=n_1s_1+n_2s_2+\cdots+n_ps_p=ns 
\end{eqnarray}
whose Jacobian is given $J=\vert ({\partial s_\ell}/{\partial t_m})\vert=\prod_\ell n_\ell^{-1}$. This allows to simplify any integral of the form 
\begin{eqnarray}\label{changing variable}
\int d\mu(s_1,\cdots,s_p)  f(s)=  \frac{1}{p!}  \prod_{\ell=1}^p \int ds_\ell f(s) = \int ds \frac{s^{p-1}}{(p-1)!} f(s)
\end{eqnarray}
for any function $f(s)$ of the variable $s$ only because $t_1 \leq t_2 \leq \cdots \leq t_p=s$. As the partition function (\ref{partition multi}) is an integral of the type (\ref{changing variable}),
we can use such a formula to simplify its expression to the form
\begin{eqnarray}\label{partition with C}
Z_p(\beta,z) = \frac{1}{p!(p-1)!} \int ds \, s^{p-1} \sum_n C_p(n)  \frac{n^p}{n!} g(n,s) z^n\exp(-\beta E(n,s))
\end{eqnarray}
with
\begin{eqnarray}
C_p(n) = \sum_{n_1+\cdots+n_p=n} \frac{n!}{\prod_{\ell=1}^p (n_\ell\, n_\ell!)}.
\end{eqnarray}
Due to the presence of this  coefficient in the series (\ref{partition with C}), we do not see how to simplify further the expression of the partition function without some approximations.  But, as for the one dimensional model,
we are exclusively interested in the behavior of the partition function at the vicinity of the Unruh temperature (where the quantum to classical phase transition occurs). Fortunately,
we can replace $C_p(n)$ by its large $n$ asymptotic to get the behavior of the partition function when $x \rightarrow 0$:  we can show that
$C_p(n) \sim p^{n+p}/n^p$ as $n \rightarrow \infty$. As a consequence, at the semi-classical limit, the grand canonical partition function behaves as
\begin{eqnarray}
Z_p(\beta,z) \sim \frac{p^p}{p!(p-1)!} \int ds \, s^{p-1} \sum_n  \frac{p^n}{n!} g(n,s) z^n\exp(-\beta E(n,s))
\end{eqnarray}
Following the same strategy as for the one color black hole model, we obtain immediately
\begin{eqnarray}
Z_p(\beta,z) \sim \frac{p^p}{p!(p-1)!} \int ds \, s^{p-4} \frac{\exp(Q_p)}{\sqrt{Q_p}}  \;\;\;\; \text{where} \;\;\; Q_p=\frac{ezp}{2} s e^{-xs}
\end{eqnarray} 
and $x=\beta \ell_p^2/2L -\pi$ as previously (\ref{variable q}). 
The expression of this asymptotic is very similar to the one color black hole model one and is given by
\begin{eqnarray}
Z_p(\beta,z) \sim Z_{p,sc}(\beta,z)=\frac{\sqrt{8\pi} \, p^p}{p!(p-1)!} \frac{x^{4-p}}{zp} \exp(\frac{pz}{2x})
\end{eqnarray}
at the vicinity of the Unruh temperature. As in the previous subsections, we now take $Z_{p,sc}$ as the grand canonical partition function.
In that case, the mean (total) number of punctures $n$, the mean (average) color $s$
and the mean area $a_H$ are given by:
\begin{eqnarray}
 s  = \frac{1}{x} \;,\;\;\;\;\;\;
 n  = \frac{pz}{2x} -1 \;,\;\;\;\;\;\;
 a_H  = 4\pi \ell_p^2 (\frac{ pz}{2 x^2}+ \frac{p-4}{x}).
\end{eqnarray}
As a conclusion, at the thermodynamical limit, the black hole area becomes macroscopic together with the total number of punctures and the average color.
Hence, the results obtained in the one color black hole are still valid when the punctures are colored with an arbitrary but finite number of colors. In particular,
the grand canonical entropy is given by (\ref{grand entropy}) and reproduces the area law with logarithmic corrections when the chemical potential is fixed to $\mu=2T_U$:
\begin{eqnarray}
S_{p,grand}(a_H)=\frac{a_H}{4\ell_p^2} + \frac{p-4}{2} \log(\frac{a_H}{\ell_p^2}) + o(\log(\frac{a_H}{\ell_p^2}) ).
\end{eqnarray}
In this formula, we see that the logarithmic corrections depend on the number of colors $p$, and
we get the coefficient $-3/2$ only for the one color black hole model. This coefficient is the one that appears in the context of real (the Barbero-Immirzi parameter is real)
black holes in Loop Quantum Gravity, and also in other approaches to black holes thermodynamics. 
As the one color black hole is the only that reproduces the ``expected" logarithmic corrections, it has something special.
In fact, it would correspond to  a  spherical symmetric quantum black hole, and maybe, implementing the spherical symmetry at the quantum level, would lead to the requirement that
all the punctures are colored by the same representation.
However, to understand properly the role of $p$, we should study the asymptotic of the partition function with an arbitrary large number of colors. We leave this study for future investigations. 

\section{Conclusion}
The content of this article is twofold. On the one hand, we construct, in a mathematically rigorous way,  the analytic continuation of the number of microstates  of a spherical black hole in loop quantum 
gravity from real values to complex values of the Barbero-Immirzi parameter $\gamma$ (in fact, our analysis is limited to $\gamma = \pm i$ which is the physically relevant case). 
On the other hand, we study the semi-classical properties of the associated statistical system.  

For a given microscopic black hole configuration (defined by a family of integers $d_\ell$), the number of black hole microstates is given by the dimension of the Chern-Simons theory Hilbert space
on a punctured 2-sphere. Hence, it is a function of the Chern-Simons level $k$ and of the integers $d_\ell$. 
The analytic continuation relies on an expression of this dimension as a complex integral along a given contour $\cal C$ of a holomorphic function. 
With this formulation at hand, the analytic continuation to $\gamma =\pm i$ becomes well defined, and consists in taking $k$ and $d_\ell$ purely imaginary, which is 
a consequence of going from $\gamma$ real to $\gamma = \pm i$. We argue that the analytic continuation is unique up to a discrete ambiguity in the choice of the contour $\cal C$.
Only one choice of contour leads to the area law at the semi-classical limit, and fortunately this choice is the simplest and somehow the most natural. 
As a consequence, this defines the number of microstates of the ``complex" black hole in Loop Quantum Gravity. 
Nonetheless, it would be very interesting to find another independent argument to justify our choice of contour. 
Such a result  is a mathematical justification of the ``naive" continuation originally in \cite{FGNP}. It is very interesting to enhance
that the intuitive analytic continuation (\cite{FGNP}) reproduced in fact the correct result not only at the leading order but also at the first subleading order, even if the regulator $\varepsilon$ (\ref{asymp naive})
was unknown.  It would be instructive to establish a link between our construction and the one proposed recently by Han in \cite{Muxin}.

Once we have constructed the number of microstates for a complex black hole, we can study its statistical and thermodynamical properties. 
We restrict ourselves to the case where the number of punctures is arbitrary but finite. 
Following \cite{Amit}, we compute the
grand canonical partition function where the energy is defined \`a la Frodden-Gosh-Perez \cite{FGP} from the point of view of 
an observer located close to the horizon, and the chemical potential $\mu$ is assumed to be identical for all the punctures. 
We show that a phase transition occurs at the Unruh temperature $T_U$. When the temperature tends to $T_U$, the system becomes classical 
in the sense that the area becomes macroscopic together with the total number of punctures and the arithmetic mean of colors.  We compute the entropy and show that it reproduces, as expected, the
area law at the semi-classical limit with some quantum correction. Wiith a very simple choice of the chemical potential, fixed to $\mu=2T_U$, the corrections are logarithmic. 

The fact that the punctures have a chemical potential means that when they disappear (in an adiabatic process for instance) they release
energy. This could be interpreted as an evidence of black hole radiation. It would be very interesting to go further and to establish a clear link between the presence of a non-vanishing
chemical potential and the new horizon radiation. Note that the logarithmic corrections come with the ``universal" factor $-3/2$ when all the punctures are colored with only one
representation. Considering a one color black hole could be interpreted as a quantization of a spherical black hole, and this might be the reason why we recover the expected logarithmic
corrections in that case only.

To finish, let us emphasize that the results of this article raise several questions and open interesting outlooks. 
The  method we have developed (the analytic continuation) is generic and could be adapted to other situations than black holes. We expect to apply them to spin foam models
and also to quantum cosmology to see how taking $\gamma=\pm i$ affects these models. Furthermore, the analytic continuation techniques could provide us with new insights to solve the reality
conditions and therefore to open a new Hamiltonian quantization of gravity starting from complex Ashtekar variables \cite{Ashtekar complex}.

\acknowledgments
We want to thank M. Geiller and A. Perez for their enthusiastic support.

\newpage

\appendix

\section*{Appendix: Analysis of the critical points}
\label{critical}
\begin{center}
\begin{figure}[h]
\includegraphics[scale=1]{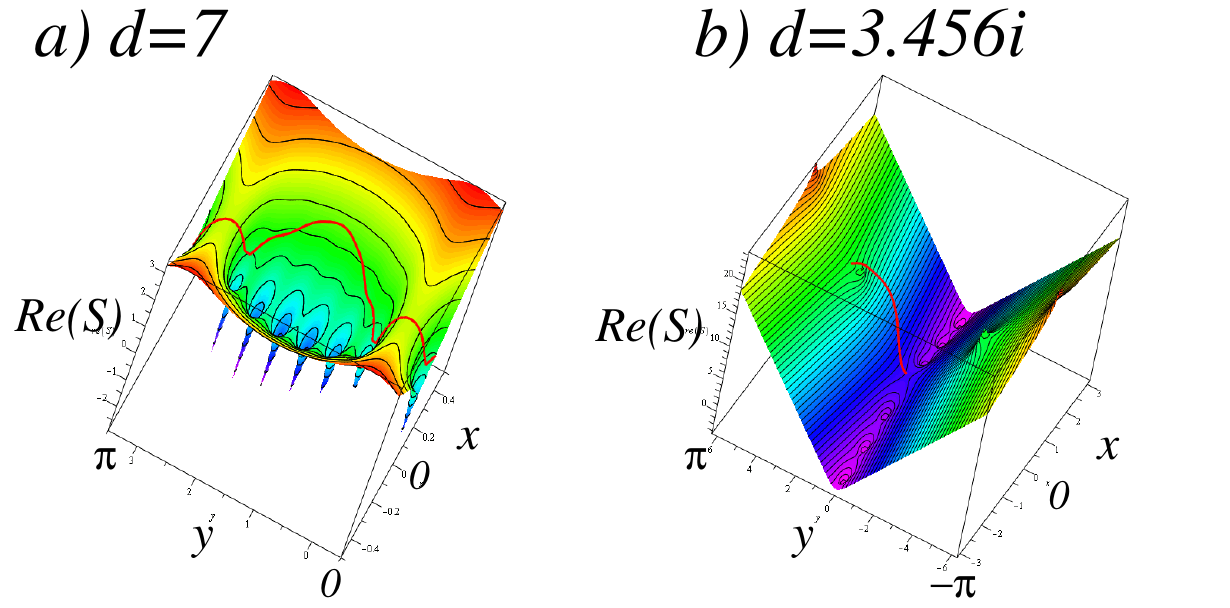}
\caption{These pictures illustrate the structure of the critical points in the complex plane of the action $\cal S$ (one puncture) for two different 
values of the representation $d$. More precisely, they represent $\text{Re}({\cal S})$ as a function of $z=x+iy$. When $d$ is an integer (here $d=7$), the critical points are located on the imaginary axe. When $d$ is imaginary,
there are two types of critical points: the ones located on the real axe, and the ones located on the imaginary axe. The contribution to the semi-classical limit of the imaginary critical points 
is clearly more important than the contribution of the real ones. All these observations are  proven in this appendix. The red line represents the component ${\cal C}_+$ of the integration contour
illustrated in the picture (\ref{contour}).}
\label{Cplus}
\end{figure}
\end{center}

This appendix contains the details of the analysis of the critical points of the action $S(z)$ 
\begin{eqnarray}\label{action app}
{\cal S}_p(z) = \sum_{\ell=1}^p \nu_\ell \log\left(\frac{\sinh(d_\ell z)}{\sinh z}\right)
\end{eqnarray}
where $\nu_\ell$ are constant and $p$ is the number of colors. This analysis is necessary to understand the large $\kappa$ expansion of the integral
\begin{eqnarray}\label{int appendix}
{\cal I}_{d_\ell}(n_\ell) = \oint_{\cal C} dz \, \mu(z) \, \exp(\kappa {\cal S}_p(z)) \;\;\;\; \text{with} \;\;\; \mu(z)=\frac{1}{i\pi}\sinh^2(z).
\end{eqnarray}
The integration contour has been illustrated in figure (\ref{contour}). The structure of the critical points in the complex plane has been illustrated in (\ref{Cplus}).

We will start with the simple case where $p=1$ and then we will generalize to an arbitrary number of punctures. 
Moreover, we will study the cases where $d_\ell$ are integer and $d_\ell=is_\ell$ are purely imaginary. 
\subsection{One color model}
We start with the case $p=1$. 
 In that case, we fix $n=\kappa$, $\nu_1=1$, $d=d_1$, we denote the integral (\ref{int appendix}) by ${\cal I}_d(n)$ and the function (\ref{action app}) by ${\cal S}$ to be coherent with the  
 notations in the core of the article.
Critical points are solutions of the equation:
\begin{eqnarray}\label{critical equation}
\tanh(dz) = d \tanh(z). 
\end{eqnarray}
The  solutions are different depending on whether $d$ is discrete or $d$ is purely imaginary. Notice that we will not study the general case
where $d$ is any complex number.

\subsubsection{The color is discrete}
Let us assume that the color $d$ is an integer ($d>1$).
First, we show that the solutions of (\ref{critical equation}) are either real or purely imaginary. To show this is indeed the case,
we decompose $z=x+iy$ in terms of its real and imaginary parts and, using
\begin{eqnarray}
\tanh(x+iy) = \frac{\sinh(2x) + i\sin(2y)}{\cosh(2x) + \cos(2y)}
\end{eqnarray}
we obtain the relation
\begin{eqnarray}\label{critical xy}
\frac{\sinh (2dx) + i \sin (2dy)}{\cosh(2dx) + \cos(2dy)} = d \, \frac{\sinh (2x) + i \sin (2y)}{\cosh(2x) + \cos(2y)}.
\end{eqnarray}
This implies necessarily the equality between the phases of the complex numbers in the r.h.s. and in the l.h.s. of the equation, i.e.:
\begin{eqnarray}
 \frac{\sinh(2dx)}{\sinh(2x)}=\frac{\sin(2dy)}{\sin(2y)} 
\end{eqnarray}
if $\sinh(2dx)$ and $\sin(2x)$ are not vanishing.
We obtain an equation of the form $f(x)=g(y)$ where $f(x)\geq d$ and $g(y) \leq d$. Furthermore, the minimum of $f$
is reached only once when $x=0$, hence the solutions of (\ref{critical equation})  are necessarily in the real axe or in the imaginary axe.
Furthermore, there is no non-zero real solutions. Therefore, all the solutions are purely imaginary i.e. $z=iy$ with
\begin{eqnarray}
\tan(dy) = d \tan(y)
\end{eqnarray}
which admits an infinity number of solutions. Trivially, if $y$ is a solution, $y+m\pi$ is solution for any $m \in \mathbb N$. Therefore, we concentrate on the interval $y \in [0,\pi]$
where the  solutions $y_m$ are labelled by $m\in\{0,\cdots,d\}$ such that:
\begin{eqnarray}
y_0=0 \;\;\;,\;\;\; y_{d}=\pi \;\;\; \text{and for} \;m=1,\cdots,d-1 \;,\;\;\; y_m \in [\frac{\pi}{2d}(2m-1), \frac{\pi}{2d}(2m+1)].
\end{eqnarray}
Notice that when $d$ is even, there is no solution for $m=d/2$. 

All these critical points could give a contribution to the large $n$ expansion of the integral (\ref{int appendix}).
However, the critical points $y_0$ and $y_d$ contribute the most because the action is maximal (when $z=i y \in i\mathbb R$)
at these points. In fact, by symmetry, these two contributions are totally identical, hence the large $n$ behavior of the integral
is obtained from the saddle point approximation at the vicinity of $y_0$, i.e.:
\begin{eqnarray}
{\cal I}_d(n) \sim \frac{2}{i\pi} \exp (n{\cal S}(0))  \int dz \,  z^2\,  \exp(n\frac{z^2}{2}{\cal S}''(0)) 
\end{eqnarray}
where 
\begin{eqnarray}
{\cal S}(0) = \log (d) \;\;\;\; \text{and} \;\;\; {\cal S}''(0)= d^2-1.
\end{eqnarray}
The integration is necessarily defined along the imaginary axe for convergence issues. Hence, we set $z=-ix$ and we obtain
\begin{eqnarray}
{\cal I}_d(n) \sim \frac{2}{\pi} d^n  \int dx \, x^2 \,  \exp(n(1-d^2)\frac{x^2}{2})  = 2 {\sqrt{\frac{2}{\pi}}} \frac{d^n}{(n(d^2-1))^{3/2}} .
\end{eqnarray}
As a consequence, the corresponding entropy admits the following large $n$ expansion:
\begin{eqnarray}
S=\log ({\cal I}_d(n)) = n \log(d) - \frac{3}{2} \log(n) + {\cal O}(1).
 \end{eqnarray}
Using the expression of the horizon area $a_H= 4\pi \ell_p^2 \gamma \sqrt{d^2-1}$, we recover the large $a_H$ expansion of the black entropy
which contains a term proportional to $a_H$ and logarithmic corrections. We need a suitable choice of $\gamma$ to reproduce the semi-classical 
area law whereas the logarithm corrections are universal and given by $-3/2 \log(a_H/\ell_p^2)$. This closes the study of the discrete case.

\subsubsection{The color is purely imaginary}
Let us now assume that the color $d=is \in i \mathbb R$ with $s>0$ for instance. We are going to show that the critical points are still either real or purely imaginary
as in the previous case. To see this is indeed the case, we write the equation (\ref{critical xy}) replacing $d$ by $is$:
\begin{eqnarray}
\frac{\sin (2sx) + i \sinh (2sy)}{\cos(2sx) + \cosh(2sy)} = s \, \frac{\sinh (2x) + i \sin (2y)}{\cosh(2x) + \cos(2y)}.
\end{eqnarray}
The r.h.s. and the l.h.s. have the same phase and then, $x$ and $y$ satisfy necessarily the condition
\begin{eqnarray}
\frac{\sin(2sx)}{\sinh(2x)}=\frac{\sinh(2sy)}{\sin(2y)}
\end{eqnarray}
if they are non-vanishing. Such a condition can be written as $f(x)=g(y)$ and it is easy to see that $\vert f \vert \leq f(0)=s$ whereas $\vert g \vert \geq g(0)=s$.
Hence, the previous requirement is fulfilled only if $x=0$ and $y=0$. As a consequence, the critical points are real of purely imaginary. 

Let us start describing the real critical points $z=x$ which are solutions of
\begin{eqnarray}
\tan(sx)=s\tanh(x).
\end{eqnarray}
 It is easy to see that there is an infinite numerable number of solutions denoted $x_m$ for $m \in \mathbb N$ such that 
 \begin{eqnarray}
 x_0=0 \;\;\;\;\text{and} \;\;\;  x_m \in [\frac{\pi}{2s}(2m-1),\frac{\pi}{2s}(2m+1)] \;\;\; \text{for}\;\; m >0.
 \end{eqnarray}
 When $z=x$ belongs to the real line, the action 
 \begin{eqnarray}
 S(x)=\log \left( \frac{\sinh(isx)}{\sinh(x)}\right) = i\frac{\pi}{2} + \log\left( \frac{\sin(sx)}{\sinh(x)}\right)
 \end{eqnarray}
 reaches its maximum at $x=0$, then it decreases exponentially. We used the principal evaluation of the logarithm with the cut on the negative real half-line.
 Therefore, among the real critical points, $x_0$ contributes the most at the semi-classical limit and the other
 contributions are suppressed. Using the saddle point approximation, we know that the $x_0$ contribution to the large $n$ expansion of the integral (\ref{int appendix}) is (up to the
 oscillating phase) governed by the evaluation of the real part of $S(x)$ at this point, namely ${\Re }(S(0))=\log(s) $.
 
 Concerning purely imaginary critical points $z=iy$, they are solutions of
 \begin{eqnarray}
 \tanh(sy) = s \tan(y)
 \end{eqnarray}
and are labelled with an integer $m$ such that $y_m = m\pi + \epsilon$. When $s \gg 1$, it is straightforward to see that $\epsilon=1/s + o(1/s)$.
We are only interested in the point $y_1$, all the others are not relevant for our choice of contour $\cal C$. 
The contribution of this point to the large $n$ asymptotic of the number of states is governed by the evaluation of the action at this point, namely
\begin{eqnarray}
{\cal S}(iy_1)= \log\left(i \frac{\sinh(sy_1)}{\sin(y_1)} \right) \simeq -i\frac{\pi}{2} + s\pi + \log(\frac{es}{2}).
\end{eqnarray}
Therefore, when $s \gg 1$, ${\cal S}(iy_1) \gg {\cal S}(0)$, which implies that the semi-classical asymptotic is totally dominated by what happens at the critical point
$z=iy_1$. We can neglect the contribution of all the other critical points. 

To finish this part, let us again point of that we have not considered the critical points $iy_p$ with $p>1$ due to our choice of contour. Another choice  would have selected another critical
point which in turn would have implied a different (larger) asymptotic behavior. The choice we made is the only one that reproduces the area law at the semi-classical limit.

\subsection{Many colors model}
\noindent The analysis of the critical points is very similar to the previous one, even if it is
a bit more involved in that case. For that reason, we will not give the details here and we will give only the relevant results. 
Now, critical points satisfy  the equation
\begin{eqnarray}
\frac{\partial {\cal S}_p}{\partial z}=\sum_{\ell=1}^p \nu_\ell \left( \frac{d_\ell}{\tanh(d_\ell z)} - \frac{1}{\tanh (z)} \right) = 0
\end{eqnarray}
As we are essentially interested in the case where $d_\ell=is_\ell$ are purely imaginary, we will not study the discrete model. Furthermore,
we know from the one color model that purely imaginary critical points have the most important contribution of  the large $\kappa$  expansion
of (\ref{int appendix}). For that reason, we will only concentrate on the purely imaginary critical points.

\noindent
As for the one color model, purely imaginary solutions are of the form $y_m=m\pi + \epsilon$ where $\epsilon \ll 1$ when the colors become large (at the semi-classical limit).
A simple calculation leads to
\begin{eqnarray}
\epsilon \simeq \frac{\sum_{\ell=1}^p \nu_\ell}{\sum_{\ell=1}^p \nu_\ell s_\ell} 
\end{eqnarray}
i.e. $\epsilon$ is the inverse of the arithmetic mean of colors. This allows to compute the large $n$ expansion of the (analytic continuation of) number of black hole
microstates given in the core of the article.

\end{document}